%% file: main.tex
\documentclass[]{interact}
\usepackage{graphicx}
\usepackage{color}
\usepackage{amsmath}
\usepackage{amssymb}
\usepackage{graphicx}
\usepackage{listings}
\usepackage{paralist}
\usepackage{makecell}
\usepackage{mathtools}
\usepackage{makecell}
\usepackage{paralist}
\usepackage{makecell}
\usepackage{graphicx}%
\usepackage{multirow}%
\usepackage{amsmath,amssymb,amsfonts}%
\usepackage{amsthm}%
\usepackage{mathrsfs}%
\usepackage[title]{appendix}%
\usepackage{xcolor}%
\usepackage{textcomp}%
\usepackage{manyfoot}%
\usepackage{booktabs}%
\usepackage{algorithm}%
\usepackage{algorithmicx}%
\usepackage{algpseudocode}%
\usepackage{listings}%

\begin{document}
	

\articletype{REVIEW}
\title{Cyber-attack TTP analysis for EPES systems}

\author{
\name{Alexios Lekidis\textsuperscript{a}\thanks{Email: alekidis@uth.gr} }
\affil{\textsuperscript{a}Department of Energy Systems \\ University of Thessaly \\ Gaiopolis Campus \\ 41500, Larissa}
}

%
%
\maketitle

\begin{abstract}
The electrical grid consists of legacy systems that were built with no security in mind. As we move towards the Industry 4.0 area though, a high-degree of automation and connectivity provides: 1) fast and flexible configuration and updates as well as 2) easier maintenance and handling of mis-configurations and operational errors. Even though considerations are present about the security implications of the Industry 4.0 \textcolor{red}{era} in the electrical grid, electricity stakeholders deem their infrastructures as secure since they are isolated and allow no external connections. However, external connections are not the only security risk for electrical utilities. The Tactics, Techniques and Procedures (TTPs) that are employed by adversaries to perform cyber-attack towards the critical Electrical Power and Energy System (EPES) infrastructures are gradually becoming highly advanced and sophisticated. In this article, we elaborate on these techniques and demonstrate them in a Power Plant of a major utility company within the Greek area. The demonstrated TTPs allow exploiting and \textcolor{red}{executing} remote commands in smart meters as well as Programmable Logic Controllers (PLCs) that are responsible for the power generator operation.
\end{abstract}


\input{intro}
\input{background}

\input{method}

\input{caseStudy}

\input{discussion}

\vspace{.2cm}
\input{conclusion}








\bibliographystyle{tfcad}
\bibliography{biblio}

\end{document}

%% file: intro.tex
\section{Introduction} \label{sec:intro}

Cyber-attacks on Industrial Control Systems (ICS) require substantial effort and sophisticated actions by adversaries due to the presence of proprietary hardware, software and communication protocols as well as the specific characteristics of each process \cite{bhamare2020cybersecurity}. Nevertheless, adversaries are attracted on targeted attacks against them since they have a massive impact to the physical world and catastrophic consequences \cite{byres2004myths}. The consequences of abnormal operation in ICS include significant risk to the health and safety of human lives and serious damage to the environment, as well as serious financial consequences such as production losses, negative impact to a nation’s economy, and compromise of proprietary information.

Electrical Power and Energy System (EPES) systems are linked to the production, transmission of electricity till its distribution in customers' households and hence belong to the ICS system category \cite{lekidis2022cyber}. Specifically, certain EPES components face a high risk from being attacked, as their replacement has great impact in the business continuity (e.g. budget loss and restoration effort). For example, the electricity production stations are more critical to a utility company than the power transmission lines. Since the focus lies on the operation of the electrical power plants to produce sufficient electricity for the existing infrastructure, security is not sufficiently tackled as a challenge \cite{kargl2014insights}. Hence, the EPES system is often not isolated, leaving open communication ports or interfaces that can be leveraged by adversaries with malicious intents. Furthermore, as the configuration of electricity components uses default or common passwords, an adversary can accordingly obtain control the functionality of the industrial process causing catastrophic consequences. 

Additionally, EPES system stakeholders believe that in the rare scenario than an adversary gains access to any of the subsystems, no harm can be caused as the underlying communications and exchanged commands cannot be interpreted.  \textcolor{red}{Nevertheless, this belief does no longer hold since proprietary communication protocols can currently be interpreted by multiple existing tools and techniques.} Specific examples are available command interpreters for 1) the Siemens S7 protocol \cite{hui2018investigating} that is used for Programmable Logic Controller (PLC) communication in electricity generation, 2) the DNP3 protocol \cite{clarke2004practical} that is used for control and data communication among Supervisory Control and Data Acquisition (SCADA) system components as well as 3) the protocol dissectors and parsers for the IEC 104 electricity substation protocol \cite{matouvsek2017description} used in Remote Terminal Unit (RTU) communication. Given the availability of protocol handling commands, the protocols are no longer restricted and proprietary. This allows unauthorized entities to not only understand the commands exchanged by in electricity systems, but also reply by sending malicious commands to them. 

The development and deployment of defense measures against cyber-attacks requires that a deep understanding of the adversary incentives behind the cyber-attacks is present. \textcolor{red}{Efforts to characterize cyber-attacks against for ICS systems have been made by the MITRE organization providing the ATT\&CK for Industrial Control Systems (ICS) framework\footnote{\small{https://attack.mitre.org/techniques/ics/}}, which depict the ICS system Tactics, Techniques and Procedures (TTPs).} In contrast to the ATT\&CK for Enterprise MITRE framework considering only security risks for IT systems, ATT\&CK for ICS is also linked with the safety risks of critical industrial systems. However, only certain TTPs of the ATT\&CK for ICS framework apply on the EPES systems. This article provides an overview of cyber-attack space for EPES systems and insights on all the relevant TTP categories. Concretely, in terms of contributions the article builds on the following:

\begin{itemize}
\item Demonstration of \textcolor{red}{a simplified EPES system architecture as well as the cyber-attack space and attack phases}.
\item A hierarchical TTP analysis for EPES systems.
\textcolor{red}{\item Application of the TTP analysis for cyber-attack. scenarios on an industrial power plant for electricity production and a smart metering infrastructure for exchanging electricity consumption measurements.  
\item Detection and mitigation measures for the industrial power plant and the smart metering scenario.}
\end{itemize}

The rest of the article is organized as follows. Section \ref{sec:background} provides an overview of EPES systems as well as the phases of a successful cyber-attack against them. Section \ref{sec:ttp} provides an overview of the TTPs that are considered in the scope of an EPES system cyber-attack. In Section \ref{sec:caseStudy} the cyber-attack TTPs are demonstrated in an industrial power plant \textcolor{red}{and a smart metering scenario. Then, Section \ref{sec:discussion} provides an overview of the lessons learned from the conducted attack scenarios, the detection and mitigation measures for them as well as the limitations of the proposed EPES system TTP analysis}. Finally, Section \ref{sec:conc} provides conclusions and perspectives for future work. 

%% file: background.tex
\section{Background} \label{sec:background}

\textcolor{red}{In this section we provide an overview of the EPES segments as well as cyber-attack spaces for EPES systems, including the distinct phases that are followed by adversaries.} 

\subsection{EPES systems}
The main segments of traditional EPES systems as depicted in Figure \ref{fig:currentGrid} are:

\begin{figure}[htbp!]
\centering
\includegraphics[width=.7\textwidth]{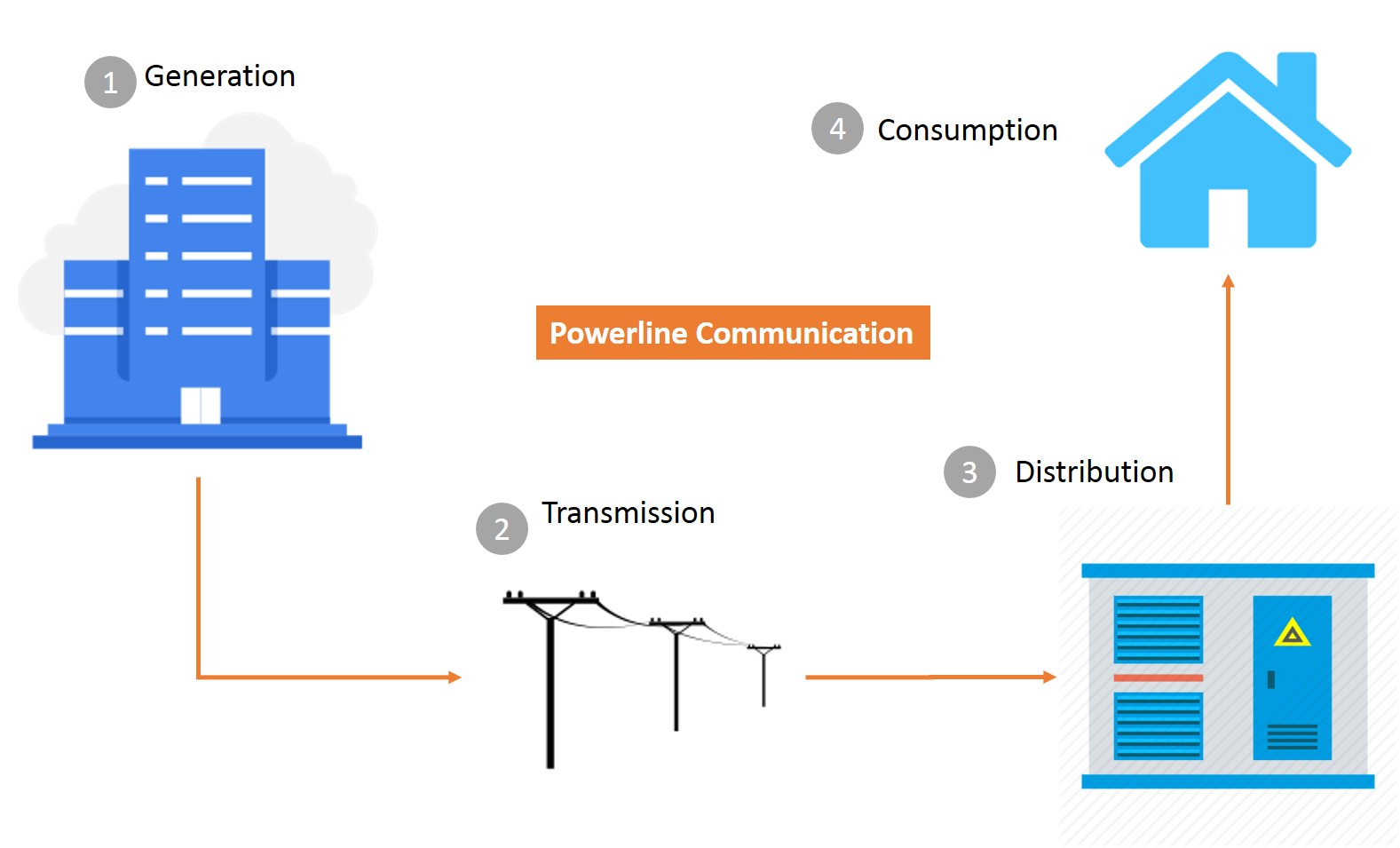}
\caption{Traditional EPES system segments}
\label{fig:currentGrid}
\end{figure}

\begin{itemize}
\item{\textbf{Generation}: This segment concerns the generation of energy from power plants of an electricity company.} 
\item{\textbf{Transmission}: The generated energy is distributed into different habitant areas (cities or neighborhoods) using power lines. The power lines are protected using switching, protection and control equipment as well as circuit breakers in order to interrupt any short circuits or overload currents that may occur on the network.}
\item{\textbf{Distribution}:  The transmitted energy is converted from high voltage power used in power lines to lower voltage power that can be used safely in residential homes and businesses. The downgrade voltage level process involves also transformers and includes data concentrators in electricity substations. The data concentrators are responsible for calculating the difference between the actual energy consumption in each individual house and compare it to an estimated consumption that is associated with historical consumption data for each house as well as with the climate conditions in the house area.}    
\item{\textbf{Consumption}: This segment involves the final use of electricity in the consumer houses or factory buildings that employ electrical appliances (e.g., meters) in order to monitor the energy consumption in the system.} 
\end{itemize}

The interactions between grid segments for electricity distribution in households and neighborhoods are mainly based on powerline communication. Powerline communication is a technology established for the transmission of electric data over the powerlines, hence it uses the existing public and private wiring for the transmission of the signals. Moreover, local Ethernet-based and serial communication technologies are used for automating and coordinating process control activities. Characteristic among them are the IEC 61850 \cite{mackiewicz2006overview} and DNP3 \cite{clarke2004practical} protocols that are used for substation automation, as well as the IEC 104 \cite{matouvsek2017description} that is used in generation.

However, the traditional grid segments are gradually becoming interconnected to exchange information that enables the transition towards the smart grid \cite{tuballa2016review}. A fundamental smart grid building block is the Advanced Metering Infrastructure (AMI) \cite{lekidis2024towards} that consists of smart meter embedded devices. AMI enables new use-cases and scenarios for the traditional electricity grid. Such scenarios concern the real-time availability of energy production and consumption data, as well as the implementation of automated customer billing scenarios based on their overall consumption. 

\begin{figure*}[htbp!]
\centering
\includegraphics[width=1\textwidth]{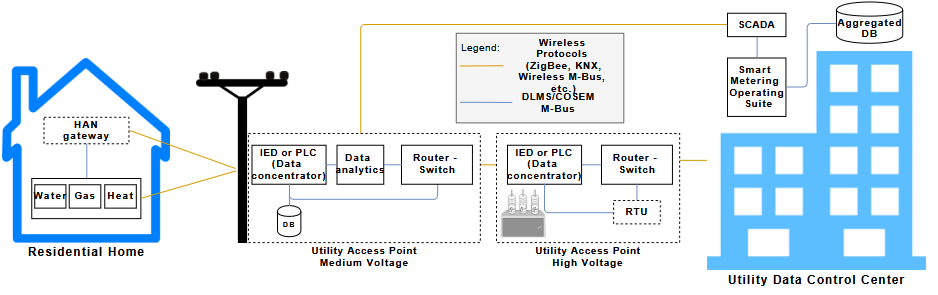}
\caption{\textcolor{red}{EPES system architecture (depicted in \cite{butun2020security})}}
\label{fig:SmartMeter}
\end{figure*}

These scenarios are enabled through intelligent devices as smart meters. Smart meters are used to collect and transmit substation measurements (e.g. current, voltage, active and reactive power) to the distribution network through cellular connection (i.e., 3G, GPRS). Through these measurements, the distribution network operator can check health status of the transformers and the network. The interactions between the EPES segments in the AMI scenario is depicted in Figure \ref{fig:SmartMeter}. 

Specifically, the \textit{Residential Home Segment} connects smart meters belonging to the same entity (home, building, facility) as well as end-user applications (Home Area Network-HAN) to a node acting as a local data collector and gateway between access and local network. This segment also has a HAN gateway were the data are centrally stored before being forwarded to the data concentrator \cite{barai2015smart}. The \textit{Utility Access Point of Medium Voltage Segment} consists of a PLC) or RTU that receives smart meter information through the wireless connection. Accordingly, it serves as a data concentrator by aggregating the data of multiple smart meters and storing them in a dedicated database. Both the actual and the estimated energy consumption are then forwarded to the next part of the EPES system through the use of a dedicated Router. 

The next segment of Figure \ref{fig:SmartMeter} is the \textit{Utility Access Point of High Voltage Segment}. This segment enables fast recovery in case of faults in the utility access point stations of medium voltage, the electricity providers use automatic reclosers as circuit breakers to restore system service. These reclosers are placed in the utility access point stations of high voltage to switch to between utility access point stations of medium voltage in case faults are spotted.  The utility access point stations of high voltage also contain a PLC as well as a Router to forward aggregated parts of the data to a centralized location of the electricity provider called utility data control center.

The last segment of Figure \ref{fig:SmartMeter} (i.e., \textit{Utility Data Control Center Segment}) contains the management system of the energy distribution company. It is defined as the backhaul network segment and has an overview of aggregated data from each consumer as well as the center for utility and customer-related services through dedicated software (i.e. Smart Metering Operating Suite in Figure \ref{fig:SmartMeter} or as AMI Headend as termed in literature \cite{sarfi2012ami}). In this segment, we can also find the SCADA of the entire system that receives all the required analytics from the dedicated component of the utility access point station of medium voltage. 
     
\subsection{Cyber-attack space}

\begin{figure*}[bthp!]
     \centering
         \includegraphics[scale=.82]{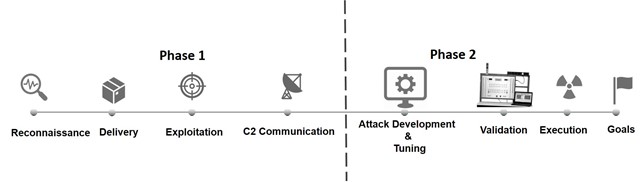}
     \caption{EPES system attack phases and associated steps \textcolor{red}{(based on \cite{jorquera2022design})}}
     \label{fig:phases}
\end{figure*}

An attack scenario in EPES systems consists of two distinct phases that are illustrated in Figure \ref{fig:phases}. The first one concerns the intrusion preparation and the second one the attack development as well as its execution on the EPES environment. Specifically, the first step has the purpose of eavesdropping information about the system, learning its behavior (Reconnaissance in Figure \ref{fig:phases}) and developing methods to evade internal perimeter protections or security mechanisms, in order to gain access to the production environment (\textit{Delivery} in Figure \ref{fig:phases}). This usually involves leveraging potential inefficiencies as entry points to the EPES system (\textit{Exploitation} in Figure \ref{fig:phases}) as well as communication to external servers to inform them of the victim, but also to be able to download attack data (Command and Control or \textit{C2 Communication} in Figure \ref{fig:phases}). As an example an adversary can penetrate the enterprise network, eavesdrop EPES connections and use potential inefficiencies as entry points to the EPES system. Inefficiencies are often exposed by the architectural complexity. On the other hand, Phase 2 seeks to cause interruption of the EPES system's normal operation. As each EPES system has its own proprietary devices, communication protocols and commands for controlling the system operation, an adversary requires this phase to cause a meaningful and reliable impact on the industrial process. Such impact is caused by developing the attack (\textit{Attack Development \& Tuning} in Figure \ref{fig:phases}) and validating it on similar or identically configured systems (\textit{Validation} in Figure \ref{fig:phases}). This is usually a long-lasting period and during its course the adversary maintains access to the system. When the attack is fully developed and validated, it still has to be executed on the industrial process (\textit{Execution} in Figure \ref{fig:phases}). Upon successful execution,  the adversary will achieve its initial goals. 

\textcolor{red}{Phase 1 steps are also present in the enterprise IT network and when compared to the 7 stages of the Cyber Kill Chain \cite{assante2015industrial} steps we can identify that the steps that are not included in Phase 1 are the \textit{Weaponization}, \textit{Installation} and \textit{Action}. The reason that these steps are not present is that they are replaced with steps that are more relevant to ICS systems and the EPES system category. Specifically, 1) the \textit{Weaponization} step is replaced with the \textit{Attack Development \& Tuning}, 2) the \textit{Installation} step is replaced with Validation and the \textit{Action} is replaced with \textit{Execution} step.}

Furthermore, as EPES devices are becoming more intelligent towards the transition to the smart grid, Phase 1 may be obsolete. This is due to the ability of compromising an Internet-facing EPES device remotely without any access on the enterprise IT network. 
Cyber-attacks on EPES systems are evolving rapidly as well as are becoming more sophisticated. Hence, the usage of only the attack phases is not sufficient for describing an adversary’s behavior and actions on an EPES system. Therefore, security analysts have conducted a study of an adversary’s behavior and actions on an EPES system, which includes the \textcolor{red}{TTPs} that are followed while the adversary is present on the EPES system. Such TTPs are part of the ATT\&CK framework for ICS by the MITRE association\footnote{https://attack.mitre.org/docs/ATTACK\_for\_ICS\_Philosophy\_March\_2020.pdf}. From these TTPs, the specific ones that are relevant for EPES systems along with associated examples are presented in Section \ref{sec:ttp}.

%% file: method.tex
\section{EPES system TTPs} \label{sec:ttp}

\begin{figure}[!htbp]
    \centering
    \includegraphics[width=0.7\columnwidth]{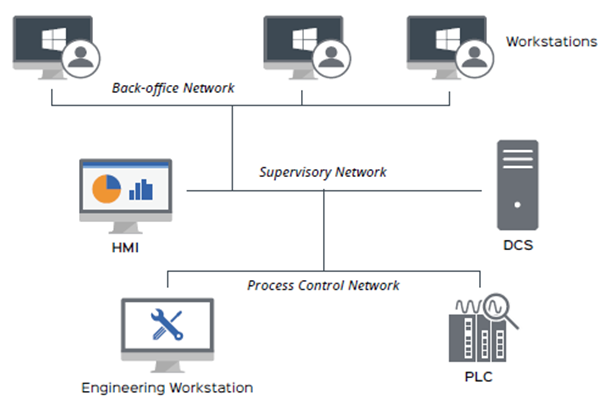}
    \caption{\textcolor{red}{Simplified EPES architecture}}
    \label{fig:epesArch}
\end{figure}

When compared to the ATTACK for ICS model, the TTPs presented in this section are a simplified version of the techniques that are applicable for EPES systems. The focus lies on tactics because they define how the adversaries are performing their actions on the EPES system. The tactics are visualized in a pyramid schematic due to the level of the EPES system that they are applicable for i.e., initial access techniques are usually applicable to the enterprise network or supervisory EPES devices whereas disruption techniques are usually applicable to field devices as industrial controllers (PLCs or RTUs). The levels of EPES systems are identical to the ICS systems, which use the Purdue Model \cite{obregon2015secure} to segment their architecture. \textcolor{red}{A simplified architecture of the EPES systems is depicted in Figure \ref{fig:epesArch}.}

\textcolor{red}{Specifically, a EPES system architecture is illustrated in Figure \ref{fig:epesArch}. The figure initially includes the \textit{Process Control Network} (Level 1 in the Purdue Model), which is composed of devices that are responsible for the configuration of critical EPES systems, such as power generators. Then, the \textit{Supervisory Network} (Level 2 in the Purdue Model) is responsible for monitoring the operation of the Level 1 components. Afterwards, the \textit{Back-office network} (Level 3 in the Purdue Model) all the IT and enterprise systems are included to provide the business operation and services of the EPES systems.}

\textcolor{red}{It is important to note here that the systems of all those layers have a lot of interconnections and may also have exposed interfaces, allowing the adversary to prepare an attack using ten different categories of tactics that are described below:}

\begin{enumerate}

\item Initial Access, defining how an adversary gains initial access on the EPES system.
\item Persistence,  defining how an adversary maintains access on the EPES system.
\item Privilege Escalation, defining how an adversary elevates its access on the EPES system.
\item Defense and Operator Evasion, defining how an adversary fools the operator or existing security mechanisms into thinking that the system behavior is normal and no attack is present.
\item Credential Access, defining how an adversary gains access to valid credentials of EPES devices.
\item Discovery, defining how an adversary locates the devices that impact the industrial process. 
\item Lateral Movement, defining how an adversary can move inside the EPES system to reach devices that impact the industrial process
\item Command \& Control, defining how an adversary communicates with external servers to receive malicious data that will cause disruption of EPES devices as well as the industrial process.
\item Execution, defining how an adversary can execute malicious code on the EPES system.
\item Disruption, defining how the adversaries can trigger physical impacts on the system (e.g., stop/degrade the process or cause catastrophic failure).

\end{enumerate}

\textcolor{red}{These categories of tactics serve as the basis for building TTP investigation workflows based on the Palantir Alerting and Detection Strategies Framework format\footnote{https://github.com/palantir/alerting-detection-strategy-framework}. Palantir Alert Alerting and Detection Strategies Framework format was chosen due to its coverage of both detection, validation and response strategies as well as the false positive indicators that can be part of detected alerts and indicators from a Security Information and Event Management (SIEM) solutions. The resulting TTP investigation workflows are provided in a dedicated dataset that is provided in \cite{dataset2025}.}

Furthermore, the techniques \textcolor{red}{associated to the tactics} refer to what actions they are performing, and finally the procedures indicate the sequence of events they are following on the EPES system. A brief summary of the techniques and procedures for EPES systems is provided in the following sections, \textcolor{red}{where each tactic is group and further explained. An important note is also that for every tactic we also provide a mapping to either the MITRE Enterprise ATT\&CK matrix\footnote{https://attack.mitre.org/tactics/enterprise/} or the MITRE ICS ATT\&CK matrix\footnote{https://attack.mitre.org/tactics/ics/}, which were considered the most relevant for EPES systems}. 

\begin{figure}[bthp!]
     \centering
         \includegraphics[scale=.72]{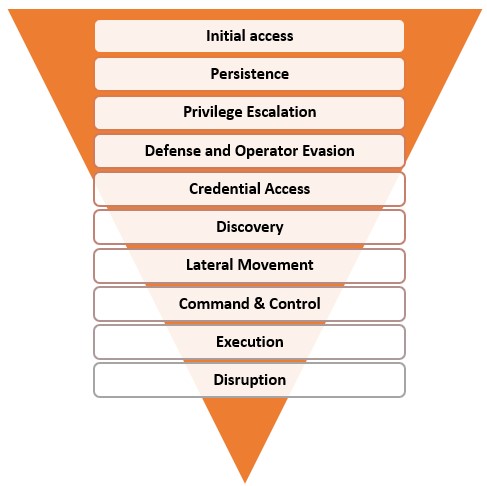}
     \caption{Cyber-attack Tactics for EPES Systems}
     \label{fig:inspection}
\end{figure}

\subsection{Initial Access}
Initial Access consists of techniques that adversaries use as entry vectors to gain an initial foothold within an EPES system \textcolor{red}{and is based on the MITRE ICS Attack Matrix \textit{TA0108}\footnote{https://attack.mitre.org/tactics/TA0108/}}. These techniques include compromising EPES devices, IT resources in the EPES system (Historian, Router, Gateway, Server \cite{assante2015industrial}) as well as external remote services and websites. Methods to compromise these devices may be the traditional spear phishing campaigns, but also remotely accessed services can be used as an entry point. Initial Access techniques may also use as entry vector supply chain personnel and sometimes may even involve gaining physical access to the EPES system. Supply chain personnel that is trusted to perform actions on the EPES system vendors, maintenance personnel, engineers, external integrators. Finally, external devices are also part of initial access techniques and include radios, controllers, removable media, that allow an initial foothold on protected EPES systems. In particular, initial access techniques include: 

\begin{itemize}
\item \textit{Spear Phishing}, is used by adversaries to gain access to industrial systems by luring targets and pretending to be legitimate system users. It relies on the established trust between the sender and the potential victims to lead them into opening an infected file that gives the adversaries control over the infected system. 
\item \textit{Supply chain compromise},  used by adversaries to gain access to the EPES systems by using supply chain personnel access. By using such techniques for the purpose of data or system compromise, adversaries can impact any component of hardware or software. 
\item \textit{Physical Access},  used by adversaries to gain access to the industrial systems by using supply chain personnel access. By using such techniques for the purpose of data or system compromise, adversaries can impact any component of hardware or software. 
\item \textit{Remote Access},  used by adversaries to gain initial access to the industrial network. Specifically, by using such techniques the adversaries can intercept remote services as VPNs, TeamViewer or Citrix that allow legitimate industrial system users (e.g. operators, supply chain personnel) access to internal industrial network resources from external locations. Remote access is given to allow remote maintenance and troubleshooting without any physical presence, especially on site that are situated in distant and harsh locations. Once initial access is obtained the adversary can eavesdrop the industrial system to gain information on the industrial network or process and accordingly plan further attack steps. Often, supply chain personnel that is unsatisfied (disgruntled employee) with the industrial manufacturer may on purpose try to cause harm by using remote access techniques. 
\item \textit{Workstation Compromise},  used by adversaries to gain initial access to the industrial network directly by compromising physical locations or stealing the identity of an operator or supply chain personnel. These locations may lack physical security handles to stop them from accessing the industrial system. Once initial access is obtained, the adversary can eavesdrop the industrial system to gain information on the industrial network or process and accordingly plan further attack steps. 
\item \textit{Data Historian Compromise},  used to gain control of a data historian to gain a foothold into the control system environment. Access to a data historian may be used to learn stored database archival and analysis information on the control system. A dual-homed data historian may provide adversaries an interface from the IT environment to the EPES environment. 
\end{itemize}

A characteristic example of initial access techniques was used in the first known attack against EPES systems using the Stuxnet malware \cite{chen2011lessons}. Stuxnet was inserted into the enterprise network using an external drive and then propagated towards the EPES devices.

\subsection{Persistence}
Often EPES systems are restarted (e.g., during maintenance, troubleshooting) adversaries may lose access to the industrial system. Losing access may also occur when legitimate credentials are modified or during a system update. Hence, adversaries should ensure that they maintain access by using persistence techniques. Such tactics include configuration changes that allow them to secure their ongoing activity and keep their foothold on EPES systems \textcolor{red}{and are based on the MITRE ICS Attack Matrix \textit{TA0110}\footnote{https://attack.mitre.org/tactics/TA0110/}}. Configuration changes may include replacing or hijacking legitimate code, firmware, and other project files, or adding startup code and downloading programs onto devices. Additionally, adversaries may trigger the execution of scheduled tasks during each restart of the EPES system.  
In particular, persistence techniques include:

\begin{itemize} 
\item \textit{External Remote Service}, used by adversaries to connect to the internal industrial network resources from external locations and maintain continuous access through an active and non-interactive service. Such techniques are used for gathering information about the industrial system and sending them to a malicious server to further plan an attack. They are used in conjunction with legitimate user credentials, in order to open remote shell to an equipment (i.e. hardware or software) of the industrial network. Prerequisite for such techniques is the availability of remote connections through for example Telnet, SSH, and VNC \cite{adepu2018assessing} on the target equipment as well as an initial network discovery to obtain information about the services that are active on the target equipment. The former usually holds, as operators and supply chain personnel often need to perform maintenance and troubleshooting on industrial without physical presence. 
\item \textit{Modify Control Logic},  used by adversaries to evade detection from system administrators and security mechanisms and therefore maintain their access to the industrial system. Modifications of control logic will include evidence of infection or certain anomalous system behavior certain as legitimate activities of the system, to lead personnel of the industrial manufacturer into thinking that the system is operating normally without any external intervention. By using such techniques the adversary ensures persistence on a potential attack by masking any existing evidence of infection as well as gains sufficient time to eavesdrop and use network discovery tactics on the industrial network, in order to interrupt the normal operation of the industrial process.
\item \textit{Module Firmware},  used by adversaries to exploit bugs in the modules that are associated with the firmware of industrial controllers, in order to obtain and accordingly maintain access to them. 
\item \textit{System Firmware},  used by adversaries to exploit bugs in the firmware of industrial controllers, in order to obtain and accordingly maintain access on them. The firmware for industrial controllers consists of the main system routines and low-level operations that use both hardware and software of the controller. To ensure persistence, an adversary may overwrite the system firmware or install malicious firmware updates, which will allow keeping the adversaries operations without any detection. To modify the system firmware the adversary has to have physical or remote access to the industrial controller, that can be obtained by other tactics such as discovery, credential access or lateral movement.
\item \textit{Valid Accounts},  used by adversaries to maintain access on industrial equipment without being detected by system administrators or security mechanisms of the industrial system. To diagnose and troubleshoot industrial operations, system administrators and operators connect remotely at regular time intervals using their credentials to the industrial system. Adversaries can perform network discovery techniques or exploit a vulnerability to obtain the credentials of a default, local or domain accounts. Default accounts are those that are built-into an OS such as Guest or Administrator account on Windows systems or default factory/provider set accounts on third-party industrial equipment. Local accounts are those configured by an organization for use by users, remote support, services, or for administration on a single system or service. Domain accounts are those managed by Active Directory Domain Services, where access and permissions are configured across systems and services that are part of that domain. Domain accounts can cover users, administrators, and services. To ensure persistence, adversaries with administrator rights can even create new users and provide them only access to specific services. 
\item \textit{Scheduled Task},  used by adversaries to maintain access to industrial equipment by scheduling a local process to execute on pre-defined time intervals. An adversary may use task scheduling to execute programs at system startup or on a scheduled basis to ensure persistence and accordingly follow lateral movement tactics, to gain access to critical systems. As a prerequisite for these techniques is the initial access to the industrial systems, that can be obtained through i.e. an external remote service or by exploiting an equipment vulnerability. Once initial access is obtained, discovery tactics need to be used, in order to capture information about the operating system and the location where the executed program will be stored. A characteristic example of using such techniques is found in the Nyetya malware \cite{rascagneres2018wasn},  which overwrites the boot sector of industrial equipment in order to successfully encrypt system information. To ensure that the system boots with the malicious boot sector and the information remains encrypted, it accordingly creates a task to reboot the system one hour after infection.
\item \textit{Hooking},  used by adversaries to modify DLLs that are associated with engineering Workstations for ensuring their persistence in the industrial system. Since engineering workstations are interfacing directly with industrial controllers, hooking allows them persistent access to the industrial process. Specifically, one type of hooking is through the Application Programming Interfaces (APIs) that are leveraged by Windows processes to perform tasks that require reusable system resources. Windows API functions are typically stored in dynamic-link libraries (DLLs) as exported functions. Another type of hooking involves redirecting calls to API functions via import address table (IAT) hooking. IAT hooking uses modifications to a process IAT, where pointers to imported API functions are stored.

\end{itemize}

A characteristic example of persistence techniques was used during the 2015 attack on the Ukrainian power grid \cite{case2016analysis}. Specifically, the adversaries gained access to the control networks of three different energy companies and developed malicious firmware for the EPES devices. The malicious firmware allowed them persistent access to the serial-to-Ethernet devices, which rendered them inoperable and severed connections between the control center and the substation. In practice this scenario would work as follows: For example, if there are three serial COM available -- 1, 2 and 3 --, the converter might be listening on the corresponding ports 20001, 20002, and 20003. If a TCP/IP connection is opened with one of these ports and held open, then the port will be unavailable for use by another party. One way the adversary could achieve this would be to initiate a TCP session with the serial to Ethernet converter at 10.0.0.1 via Telnet on serial port 1 with the following command: telnet 10.0.0.1 20001.

\subsection{Privilege Escalation}
Privilege Escalation consists of techniques that adversaries use to gain higher-level permissions on a system or network \textcolor{red}{and is based on the MITRE ICS Attack Matrix \textit{TA0110}\footnote{https://attack.mitre.org/tactics/TA0110/}}. Adversaries can often enter and explore a network with unprivileged access but require elevated permissions to follow through on their objectives. Techniques that are included within privilege escalation usually involve exploitation of vulnerabilities that allow to take advantage of system weaknesses, misconfigurations, and vulnerabilities. Additionally, privilege escalation may be triggered using valid accounts, such as the root in Linux systems, local administrator in Windows systems and user accounts with access to EPES devices that perform critical functionalities. These techniques often overlap with Persistence techniques, as OS features that let an adversary persist can execute in an elevated context.
In particular, privilege escalation techniques include: 

\begin{itemize} 
\item \textit{Exploitation of Vulnerability},  used by adversaries to gain access to administrative or root accounts of industrial systems that would allow them to execute arbitrary commands on the industrial process. Such techniques are successful if an adversary has access to the industrial network and can use credential access tactics to first obtain credentials of legitimate user accounts and then exploit vulnerabilities of industrial equipment as Workstations to elevate the obtained privileges. Elevated privileges are important to bypass safety mechanisms of industrial controllers that allow critical commands to industrial controllers only from privileged users as network administrators or experienced operators.  
\item \textit{Valid Accounts},  which do not only allow adversaries to gain access to an industrial network, but also to escalate their access to a higher permissions level for orchestrating and performing actions on the industrial process. Elevated privileges can also be used to bypass access controls or any security mechanisms that are set within the industrial manufacturer. In order to perform privilege escalation with valid account techniques, a prerequisite of an adversary is to have at least normal-level access to the target system. This can be obtained through other tactics, as Credential Access.
\end{itemize}

A characteristic example of such techniques lies with Siemens SINUMERIK controllers that were discovered (CVE-2018-11462) \footnote{https://nvd.nist.gov/vuln/detail/CVE-2018-11462} to have improper control of privileges, that allows adversaries to escalate privileges to an elevated user account, but not to root. To exploit this vulnerability, an adversary would only require network access to the SINUMERIK controllers and neither specific privileges nor any user interaction.

\subsection{Defense and Operator Evasion}
Adversaries use such techniques to avoid detection from operators or security controls/system protection mechanisms during the course of their attack. \textcolor{red}{These techniques are based on the MITRE ICS Attack Matrix \textit{TA0110}\footnote{https://attack.mitre.org/tactics/TA0103/} (i.e., Evasion) tactic. Moreover, they are used} for evasion include removal of indicators of compromise, spoofing communications and reporting, and exploiting software vulnerabilities. Adversaries may also leverage and abuse trusted devices and processes to hide their activity, possibly by masquerading as master devices or native software. Techniques of defense and operator evasion have an overlap with Execution and Disruption techniques, however their main difference is that defense and operator evasion does not aim in causing impact on the industrial processes or EPES. Additionally, defense and operator evasion may also vary depending on whether the target of evasion is human or technological in nature, such as security controls. Examples of these techniques include the deletion or modification of files, logs, alarm settings, device displays and reports of supervisory devices of EPES systems. Others are modifying directly the EPES devices that are inputs/outputs or the operational state of industrial controllers that are communicating directly with the industrial process. In particular, defense and operator evasion techniques include: 

\begin{itemize} 
\item \textit{Alternate Modes of Operation},  used by adversaries to mask a potential attack to industrial equipment. To perform such techniques, access (e.g. remote or physical) to the industrial controllers (e.g., PLCs, RTUs) or to perform a Man-In-The-Middle (MITM) attack between a Workstation or a Human-Machine Interface (HMI) and an industrial controller. As a technique, process adversaries may forge or replay messages showing a normal controller operation, whereas in reality the controller might be in a different operating mode and possibly dis-functional. 
\item \textit{Block Communication Port},  used by adversaries to avoid detection by operators while executing targeted attacks on industrial equipment as Workstations. These techniques are successful when the adversary gains access to the industrial equipment either physically or remotely by credential access tactics. Additionally, adequate system knowledge through a network discovery or eavesdropping phase is also required to know which ports have direct impact on the industrial process. 
After a successful blocking of communication ports, an adversary can replay eavesdropped messages and possibly alter legitimate commands to mask the ongoing attack. 
\item \textit{Block Reporting Message}, used by adversaries to execute targeted attacks on industrial systems by simultaneously avoiding detection from operators or security mechanisms. SCADA systems receive periodically as well as can also request information about the functionality of the industrial process from Workstations, PLCs and RTUs. Adversaries can perform MITM attacks, in order to intercept messages that are reporting indicators of anomalous system behavior that is caused by their attack.    
\item \textit{Exploitation of Vulnerability},  used by adversaries to bypass security mechanisms and execute their code on the industrial systems. Such techniques allow for remote infiltration of the enterprise network or even directly the industrial network by leveraging a mis-configuration of a security (e.g., firewall, endpoint protection) or industrial equipment as the SCADA and Workstation systems.  A characteristic example, depicted by Gartner \cite{brazil2014security}, is that firewall breaches are due to mis-configured firewalls and no security flaws. Instead of segmenting networks and preventing unauthorized access to critical industrial equipment, enterprises usually have no segmentation and allow all communications between the business and the industrial network. Another example lies in the vulnerability that was exploited by the Triton malware and which disabled the firmware RAM/ROM consistency check that allowed to read and write malicious data into the memory of Triconex controllers without any verification. 
\item \textit{File Deletion}, used by adversaries to mask their attacks on the industrial process by deleting diagnostic files that are stored in the SCADA or the Historian systems. If adversaries manage to successfully delete files as a part of a cleanup process, operators and security mechanisms cannot locate traces that can be used as evidence of an attack. A characteristic example of these techniques was the LockerGoga ransomware \cite{adamov2019analysis} that deleted all backups, which would have allowed the restoration of files that were encrypted. 
\item \textit{Masquerading},  used by adversaries to disguise a malicious application or executable as another file, to avoid operator and engineer suspicion. Possible disguises of these masquerading files can include commonly found programs, expected vendor executables and configuration files, and other commonplace application and naming conventions. By impersonating expected and vendor-relevant files and applications, operators and engineers may not notice the presence of the underlying malicious content and possibly end up running those masquerading as legitimate functions. Applications and other files commonly found on Windows systems or in engineering workstations have been impersonated before. This can be as simple as renaming a file to effectively disguise it in the industrial environment.
\item \textit{Modify Event Log}, used by adversaries to remove events and warnings from industrial systems to avoid suspicion of operators for indicators of anomalous behavior that is caused from ongoing attacks. Event logs are accessed frequently by operators to keep track of operations that are performed on Workstations, PLCs or RTUs i.e. patches, updates, maintenance schedules. Due to the resource constraints of PLCs and RTUs event logs are stored in Workstations as well as SCADA systems. Hence, to perform such techniques, adversaries need to have privileged access on the Workstation or SCADA systems to modify or even delete logged events from the associated files. In such case, a possible attack will have no evidence for forensics and will be considered as a possible failure. An example is by considering a water tank that is operating automatically without any human intervention. If an adversary modifies the tank to be operating manually with human intervention, it can then overflow the water from the tank. Changing the tank operation will generate a message from the Workstation to the SCADA. If this message is removed from the event log, a potential overflow of the tank will lead the operator into thinking that a failure occurred instead of an attack.   
\item \textit{Modify HMI/Historian Reporting},  used by adversaries to alter stored diagnostic reports of industrial equipment to avoid suspicion of operators for indicators of anomalous behavior that is caused from ongoing attacks. HMIs provide a graphical interface where operators can access low-level information about the industrial equipment. HMIs provide important system information such as metrics, alerts, current functions, and other pertinent system data. HMI information are also backed up as reports in historians, that are databases with critical system data, allowing operators to industrial system administrators to have an overview of performed activities on the industrial system. Adversaries may modify such reports to remove possible traces of their malicious activity on the industrial system.
\item \textit{Modify I/O Image},  used by adversaries to mask their attacks on the industrial process by deceiving operators into thinking that the industrial system is exhibiting normal behavior. This is achieved by modifying the I/O process image, which indicates the memory area where industrial controller i.e. PLC data from the I/O modules are located and can be copied to/from. Hence, these techniques allow the adversary to change the mapping between I/O modules and memory areas by simultaneously keeping the outputs to the previous values. This will evade any suspicion from operators, but will also introduce errors on the industrial process at a later stage. Modify I/O image techniques require advanced knowledge of the industrial controller that can be obtained with long-lasting eavesdropping as well as discovery tactics.    
\item \textit{Modify Physical Device Display},  used by adversaries to mask their attacks on the industrial process by deceiving operators into thinking that the industrial system is exhibiting normal behavior. Physical Device Displays are used for high-level monitoring and diagnostics of the industrial process. Systems that include such displays are the HMI and the SCADA systems. The adversary can execute code or replay legitimate messages to illustrate on these display that everything is operating normally, whilst manipulating the industrial controller logic or executing commands on the industrial process.
\item \textit{Modify System Settings},  used by adversaries to hide malicious activities that would raise suspicion to operators or system administrators. System settings in an industrial system may consist of warnings and indications for abnormal activities in the industrial process, e.g. extremely high water level in a tank. Adversaries may change or even disable such settings, in order to cause catastrophic consequences on the industrial process. Industrial equipment that is associated with such settings is the SCADA and the HMI. Additionally, adversaries may also disable any security mechanism would alert the system administrators. In order for these techniques to be successful, the adversary requires knowledge of the industrial system and in particular of the processes that SCADA and HMI systems use to generate these warnings and indications of abnormal activities.   
\item \textit{Rootkit},  used by adversaries to hide the existence of malware by intercepting and modifying operating system API calls that supply system information. Rootkits may be used in industrial equipment as Workstations or PLCs, to hide the presence of programs, files, network connections, services, drivers, and other system components. A characteristic example of such techniques is the rootkit used by Stuxnet \cite{falliere2011w32}, allowing to hide all malicious files and processes to evade detection. Specifically, Stuxnet injects the rootkit onto the PLC and Step 7 software, modifies the code, and sends commands to the PLC while displaying system information that indicates a normal operation to the end user. 
\item \textit{Modify alarm settings},  used by adversaries, to prevent alerts that may inform operators of their presence or to prevent responses to dangerous and unintended scenarios. Reporting messages are a standard part of data acquisition in control systems. Reporting messages are used as a way to transmit system state information and acknowledgments that specific actions have occurred. These messages provide vital information for the management of a physical process, and keep operators, engineers, and administrators aware of the state of system devices and physical processes. If an adversary is able to change the reporting settings, certain events could be prevented from being reported. This type of modification can also prevent operators or devices from performing actions to keep the system in a safe state. If critical reporting messages cannot trigger these actions, then significant impact could occur.
In industrial systems, the adversary may have to use Modify alarm settings or contend with multiple alarms and/or alarm propagation to achieve a specific goal to evade detection or prevent intended responses from occurring. Methods of suppression often rely on modification of alarm settings, such as modifying in memory code to fixed values or tampering with assembly level instruction code.
\end{itemize} 

A characteristic example of defense and operator evasion techniques is the Irongate malware \cite{firoozjaei2022evaluation}, which recorded five seconds of 'normal' traffic from a PLC to the user interface and replayed it, while simultaneously it continued sending different data back to the PLC to change its operating mode. This allows an adversary to alter a controlled process unbeknownst to process operators.

\subsection{Credential access} 
    Credentials access techniques are used by adversaries for stealing legitimate credentials as account names and passwords, in order to gain remote access to EPES devices \textcolor{red}{and are based on the MITRE Enterprise ATT\&CK Matrix\footnote{https://attack.mitre.org/tactics/TA0006/} tactic.} Using legitimate credentials can give adversaries access to systems, make them harder to detect, and provide the opportunity to create more accounts to help achieve their goals. Techniques used to obtain credentials include brute forcing combinations of username and password pairs, as well as credential dumping and network sniffing techniques. Through a successful login to an EPES device, adversaries can then use further tactics (e.g., lateral movement, execution, disruption) to interrupt the normal operation of the industrial process. In particular, credential access techniques include: 
\begin{itemize} 
\item \textit{Brute Force},  used by adversaries in order to obtain the current credentials of an industrial device. When such techniques are successful, the adversary can obtain root access to industrial devices. A characteristic example of using such technique is the S7 Password Offline Bruteforce Tool of the Scadastrangelove group\footnote{https://pastebin.com/0G9Q2k6y}.
\item \textit{Credential Dumping},  used by adversaries in order to obtain from the operating system and software, the credentials of industrial equipment in the form of a hash or a clear text password. Industrial equipment that is relevant for such techniques is Workstations, Human-Machine-Interfaces (HMIs) and  SCADA systems. When such techniques are successful, the adversary can obtain access to such equipment and further use lateral movement tactics to cause direct impact on the industrial process. A characteristic tool that can be used as a credential dumper is Mimikatz\footnote{https://github.com/gentilkiwi/mimikatz}. Mimikatz is capable of obtaining plain-text Windows account logins and passwords, along with many other features that make it useful for testing the security of networks. When advanced industrial controllers are used, an equivalent tool of Mimikatz for the Linux operating system is used called MimiPenguin\footnote{https://attack.mitre.org/software/S0179/}. MimiPenguin tries to harvest credentials from the /proc file system on Linux, which contains important information for the state of the running operating system.
\item \textit{Default Credentials},  used by adversaries to gain remote access to industrial equipment. Industrial manufacturers buy industrial equipment (i.e. hardware or software) from third-party vendors (e.g. Siemens) that is set up with default passwords. To avoid losing the password, that will require to reset the equipment from the beginning, manufacturers usually maintain the default password. However, vendors mention the password in manuals or equipment documentation that is either available online or used in other equipment of the same vendor. This allows adversaries to obtain this password and try to log in remotely to the industrial equipment. 
\item \textit{Network Sniffing},  used by adversaries to obtain valid user credentials that can be later used to log in successfully to industrial equipment. Industrial manufacturers usually employ no encryption schemes when they exchange data through their proprietary network protocols, and hence usually transmit credentials in clear text. A characteristic example of using network sniffing techniques is with the Siemens S7 proprietary protocol\footnote{https://www.us-cert.gov/ics/alerts/ICS-ALERT-11-204-01B}. As a prerequisite for the usage of network sniffing techniques is the support for the proprietary industrial protocols and/or communication profiles, in order to interpret the exchanged credential messages and intercept the user credentials in them. Examples of open-source tools providing network sniffing techniques for EPES networks are Wireshark\footnote{https://www.wireshark.org/}, Zeek\footnote{https://zeek.org/}, Suricata\footnote{https://suricata.io/} as well as many other proprietary network analysis tools.
\end{itemize} 

A characteristic example of credential access techniques is with EPES devices that are set up with default credentials. Specifically, Industrial manufacturers buy industrial equipment (i.e., hardware or software) from third-party vendors (e.g., Siemens) that is set up with default passwords. To avoid losing the password, that will require to reset the equipment from the beginning, manufacturers usually maintain the default password. However, often device users mention the password in manuals or equipment documentation that is either available online or used in other equipment of the same vendor. This allows adversaries to obtain this password and try to log in remotely to the EPES devices. A list of default username/password pairs are available by the Scadastrangelove group \cite{caselli2016security}. Through a successful login to the EPES devices, adversaries can then use further tactics (e.g. lateral movement, execution, disruption) to interrupt the normal operation of the industrial process.

\subsection{Discovery}
Discovery consists of techniques that adversaries use to gain knowledge about the internal EPES system network, control system devices, and how their processes interact \textcolor{red}{and is based on the MITRE ICS Attack Matrix \textit{TA0102}\footnote{https://attack.mitre.org/tactics/TA0102/} tactic}. These techniques help adversaries observe the environment and determine the next steps of their attack, such as the use of Lateral Movement techniques. Discovery techniques are often an act of progression into the EPES system and enable the adversaries to orient themselves before deciding how to act. They include the discovery of EPES system networks, locations, active network services, serial connections as well as allow adversaries to move even further into discovering industrial control devices, their input/output interfaces and the control process that is running on the EPES system. In particular, discovery techniques include: 

\begin{itemize}
\item \textit{Control Process},  used by adversaries to discover the normal operation of the industrial process that is carried out by the industrial manufacturer. Such techniques require sufficient eavesdropping and analysis of all the information exchanged in the network. Hence, an adversary may require a long time duration for such tactics to be successful. During Stuxnet \cite{falliere2011w32} these techniques were used as adversaries infiltrated the enterprise network one year before it started malicious actions on the industrial network.
\item \textit{I/O Module Enumeration},  used by adversaries to discover the input and output values of PLCs, which indicate the values of normal operation for the industrial process. Module enumeration is performed by using dedicated protocol messages that are manufacturer specific and are sent in clear text without any authentication. No authentication allows the Workstations or HMIs to read PLC inputs and outputs, in order to perform diagnostics and troubleshooting. However, to employ such techniques an adversary needs to be present in the industrial network, in order to issue such the required protocol messages pretending to be a legitimate Workstation or HMI. In Siemens S7 PLCs the read SZL lists contain I/O information. By leveraging I/O Module Enumeration techniques, an adversary can learn the industrial process and later issue commands that will interrupt its normal operation. 
\item \textit{Location Identification},  used by adversaries to discover the exact placement (i.e., physical location) of industrial equipment in the industrial network. Such techniques aim at identifying the network structure, its connections with other networks and the details of its subnets. As a next step, the adversary can obtain their IP address, the subnet and LAN they belong to, in order to finally determine where they are located. Certain application protocols (e.g. Modbus, Profinet) also include location information in the exchanged data. An example tool that can be used for location identification techniques is Kamerka\footnote{https://github.com/woj-ciech/Kamerka-GUI}, which is able to search for industrial equipment in a specific country. Kamerka extends Shodan ICS\footnote{https://www.shodan.io/explore/category/industrial-control-systems}, that performs crawling, IP lookups, searching, data streaming for industrial networks. 
\item \textit{Network Connection Enumeration},  used by adversaries to detect active or dormant equipment in an industrial network. Such techniques aid adversaries in: locating critical devices (Lateral Movement/Exploitation of Remote Services) and extracting information about the vendor as well as the firmware of industrial controllers. Combined with network service enumeration, these techniques are the first and most common step used by malicious actors to launch attacks against EPES devices, as well as an early attack indicator for defenders. Network connections can be retrieved from a network element (e.g. router, switch) and specifically by using SNMP (e.g. through snmpwalk or snmpget commands) \cite{stallings1993snmp}. 
\item \textit{Network Service Scanning},  used by adversaries to detect open ports and services running on EPES network equipment. Such techniques aid adversaries in: 1) locating critical devices (Lateral Movement/Exploitation of Remote Services) and extracting information about the vendor as well as the firmware of industrial controllers as well as 2) locating open ports and services that are running on the industrial equipment and can be remotely accessed as an initial access tactic for an attack. Thus, this is the first and most common step used by malicious actors to launch attacks against industrial equipment, as well as an early attack indicator for defenders. PLCScan tool\footnote{https://github.com/meeas/plcscan} is an open-source tool used for network service scanning.
\item \textit{Network Sniffing},  used by adversaries to eavesdrop the communications and data exchanged in an EPES network. As a prerequisite for such techniques, an adversary needs to have already obtained initial access to the enterprise or industrial network as well as the ability of real-time monitoring of all its connections. As a favorable scenario, adversaries may aim at accessing to the mirroring port of a network element (e.g., router, gateway) or specifically the Switched Port ANalyzer (SPAN) port on a switch, which will allow them to capture all the network communications on any other port of the network element. Example tools for network sniffing are Wireshark and tcpdump\footnote{https://www.tcpdump.org/}. Network sniffing techniques are passive and do not provide evidence or trace on the network, hence an adversary can adversary cannot cause any harm with them on the network but also evades detection from operators or system administrators. Therefore, they are used in the initial steps of an attack. Finally, network sniffing can also be performed by unintentional actors (e.g. operators, system administrators) for maintenance and troubleshooting scenarios. 
\item \textit{Role Identification},  used by adversaries to discover the role that industrial equipment has in the system by interpreting network information. As a prerequisite for such techniques, an adversary needs to have already obtained initial access to the industrial system, as well as the ability either poll or interpret the industrial protocol and messages that the equipment transmits. Often, for intercepting the role of industrial equipment, contextual information about its characteristics is required, such as the model, firmware, and  back-plane  connections  (non-IP  equipment connected  to  industrial  systems). Certain application layer protocols (e.g., Modbus \cite{swales1999open}, Profinet \cite{feld2004profinet}) provide a lot of contextual information to facilitate role identification. An example role identification tool is the p0f tool\footnote{http://lcamtuf.coredump.cx/p0f3/}.
\item \textit{Serial Connection Enumeration},  used by adversaries to discover the serial connections that are active for an industrial network. Since industrial networks usually contain legacy technologies, there are often serial connections between industrial controllers and/or field equipment, which is directly connected to the physical process (e.g., motors, turbines). Example EPES devices using serial connections are the Modbus RTU that is based on the RS485 standard \cite{axelson1999designing}. Adversaries may use such techniques to gain knowledge of the physical process. Enumerating serial connections requires adequate knowledge of the serial network protocols that are used, as well as access to the equipment that is connected over serial wires or through a USB connection with the target serial controller. By enumerating serial connections, an adversary can intercept the port, baud rate, data bits, parity, stop bits, and flow control of industrial equipment. Moreover, through the discovery of serial connections an adversary can use connection-less commands to serial equipment, in order to interrupt the normal process operation. Enumerating serial connections requires adequate knowledge of the serial network protocols that are used, as well as access to the equipment that is connected over serial wires with the target serial controller. Connection enumeration can be performed through the libserialport library\footnote{https://sigrok.org/wiki/Libserialport}.
\item \textit{Control Device Discovery},  used by adversaries to discover industrial controllers that are involved in the normal operation of the industrial process, which is carried out by the industrial manufacturer. Such techniques require sufficient eavesdropping and analysis of all the information exchanged in the network. Hence, an adversary may require a long time duration for such tactics to be successful. In this category, we can find the Digital Bond's ICS Enumeration Tools\footnote{https://github.com/digitalbond/Redpoint}, that can be used to retrieve industrial controller information. 
\end{itemize}

A characteristic example of using discovery techniques is through the nmap library. Nmap includes a scripting library with scripts for EPES devices, such as Siemens and Rockwell PLCs. The scripts can obtain Device Type, Vendor ID, Product name, Serial Number, Product code, Revision Number, status, state, as well as the Device IP. Discovery techniques as using nmap were employed by Stuxnet, that infiltrated the enterprise network one year before it started malicious actions on the EPES system network.

\subsection{Lateral movement}
Lateral Movement techniques allow adversaries to enter and control remote systems on a EPES network \textcolor{red}{as well as are based on the MITRE ICS Attack Matrix \textit{TA0109}\footnote{https://attack.mitre.org/tactics/TA0109/} tactic}. These techniques leverage default credentials, known accounts, and vulnerable services,  to pivot to their different devices of the EPES environment. Using such techniques, adversaries can position themselves where they want to be or think they should be. Following through on their primary objective often requires Discovery of the network to develop awareness of unique EPES devices and processes, in order to find their target and subsequently gain access to it. Reaching this objective often involves pivoting through multiple systems, devices, and accounts. 

Adversaries may install their own remote tools to accomplish Lateral Movement or leverage default tools, programs, and manufacturer set or other legitimate credentials native to the network, which may be stealthier. Additionally, they can use local services of the EPES system or can even exploit a vulnerability (e.g., use of insecure protocols or services) to move across the system. Finally, often for moving laterally across the EPES system, an adversary may require legitimate user credentials or valid accounts. In particular, lateral movement techniques include: 

\begin{itemize}
\item \textit{Default Credentials}, used by adversaries to move laterally across devices of the industrial environment using legitimate credentials upon authentication. ICS manufacturers usually setup industrial devices with a default password that is not updated once the device is deployed in the industrial environment. These passwords are factory set and are often easy to guess or are changed infrequently, which creates additional security risks. The list of default passwords for each ICS device vendor is found in\footnote{\small{https://github.com/scadastrangelove/SCADAPASS}}. Once an adversary gains access either on the enterprise or industrial network, a common strategy is to identify all connected devices and try to log in to them by trying all combinations of default credentials. By iterative lateral movement tactics using default credential techniques, an adversary can often gain direct access in the industrial process.     
\item \textit{Exploitation of Vulnerability}, used by adversaries to take advantage of programming errors in a program, service, or within the operating system software or kernel itself to gain access to critical industrial devices. By using lateral movement techniques, adversaries are able to access the critical systems of oil refineries, city infrastructure objects and electrical distribution network facilities.    
\item \textit{External Remote Service}, used by adversaries to gain access to industrial systems remotely. Such techniques are feasible when a service is running on a port of an industrial device and can be accessed remotely through web services. Remote services are used when operators and supply chain personnel need to perform maintenance and troubleshooting to industrial equipment that is present in distant or harsh locations and can be accessed rarely. Usually, such services require authentication through valid credentials that can be obtained by either credential access or discovery techniques. Often, enterprises use the credentials that are set up by industrial equipment vendors, which allows the adversary to obtain access to the industrial system with minimum effort.
\item \textit{Internal Service}, used by adversaries to gain access to industrial systems through lateral movement from infected systems of the enterprise network. Such techniques are feasible when a service is running on a port of an industrial device and can be accessed directly from the enterprise network. This is due to the absence of network segmentation that is usually spotted in enterprises with industrial environments. Direct connections to the services of the industrial environment allow automated updating and maintenance functionalities, that do not require the physical presence of operators or supply chain personnel on the industrial site. This indicates that internal service techniques can be also employed by legitimate personnel.
\item \textit{Valid Accounts}, used by adversaries to gain access to industrial systems by impersonating a  legitimate user. Such techniques are feasible if a legitimate user identity or corresponding credentials as well as public and private keys have been stolen or eavesdropped. Adversaries can use three categories of accounts: default, local, and domain accounts. Default accounts are not limited to Guest and Administrator on client machines, but they also include accounts that are preset for industrial equipment and are not modified by operators or supply chain personnel when it is installed on the industrial site.
\end{itemize}

A characteristic example of using Lateral Movement techniques are  the enterprise network exploits that propagated to the industrial network, such as the exploits of vulnerability SMBv1 targeting EternalBlue or MS17-010 \footnote{https://learn.microsoft.com/en-us/security-updates/securitybulletins/2017/ms17-010}. The ransomware that exploited successfully this vulnerability are Bad Rabbit \cite{alotaibi2021sdn}, NotPetya \cite{fayi2018petya} and WannaCry \cite{mohurle2017brief}. The exploitation used computers (servers and even workstations) that are connected to several subnets inside the organization’s perimeter at the same time. The ransomware used port 445 that was available both for the enterprise and the industrial network, as it was required for migrating historian data to business intelligence systems.

\subsection{Command and Control}
Command and Control techniques are used by adversaries when they are trying to control compromised systems, controllers by, i.e., exchanging commands and data with external servers \textcolor{red}{and are based on MITRE ICS Attack Matrix \textit{TA0101}\footnote{https://attack.mitre.org/tactics/TA0101/}}. Usually, adversaries combine these techniques with defense and operator evasion techniques, in order to mimic expected network traffic to avoid detection and suspicion. For example, an adversary can use a common port for the enterprise network or the EPES system to communicate externally, such as the tcp/80 (HTTP), tcp/443 (HTTPS), tcp/25 (SMTP) and tcp/53, udp/53 (DNS) enterprise ports. Corresponding EPES system ports include the ones mentioned in Table \ref{tab:protocols}.

\begin{table}[htbp!]
  \centering
    \resizebox{8.5cm}{!} {
    \begin{tabular} {|c||c|}
        \hline
        \textbf{EPES Protocol} & \textbf{Communication Port} \\
    \hline
Siemens Step 7 & TCP/102 \\
\hline
IEC 61850 & TCP/102 \\
\hline
Modbus & TCP/502 \\
\hline
BACnet/IP & UDP/47808  \\
\hline
Ethernet/IP & TCP/44818 - UDP/44818 - UDP/2222 \\
\hline
DNP3 & TCP/20000 - UDP/20000 \\
\hline
PROFINET & TCP/34962 - - UDP/34962 \\
\hline
EtherCAT & UDP/34980 \\ 
\hline
IEC 104 & TCP/2404 \\
\hline
Niagara Fox & TCP/1911 - TCP/4911 \\
\hline
ProConOS	 & 	TCP/20547 - UDP/20547 \\
\hline 
PCWorx & TCP/1962 \\
\hline 
Moxa & 	TCP/4800 \\
\hline 
OMRON FINS & TCP/9600 - UDP/9600 \\
\hline 
Melsoft & UDP/5006 \\
\hline
CoDeSyS & TCP/1200 - TCP/2455 \\
\hline
\end{tabular}}
\vspace{.2cm}
\caption{EPES protocols and associated communication ports} 
  \label{tab:protocols}%
\end{table}

Other scenarios where these techniques are used include also uncommon ports for communicating to external servers allowing adversaries to avoid having the communication intercepted by firewalls or bypass other EPES protection systems, such as proxies. An example here is the usage of port 1502 by the Tristation protocol \cite{di2018triton} that is operating as an application protocol on top of UDP data frame. 
Command and Control may be established to varying degrees of stealth, often depending on the victim’s network structure and defenses. In particular, command and control techniques include: 

\begin{itemize}
\item \textit{Commonly Used Port}, used by adversaries for command and control communication to external malicious server. Common ports are usually unfiltered by the firewall, to allow inbound and outbound communication for enterprise services (e.g. email) or industrial protocols (e.g., Siemens Step 7, Modbus). Examples of such ports are tcp/80 (HTTP), tcp/443 (HTTPS), tcp/25 (SMTP) and tcp/53, udp/53 (DNS). During the Trisis attack in the Saudi petrochemical plant \cite{di2018triton}, port 443 (HTTPS) was used to connect to an external server, in order to download the binaries that would replace the engineering workstation executable.
\item \textit{Connection Proxy}, used by adversaries to direct network traffic between systems or act as an intermediary for network communications to a command and control server. A characteristic example is with the Stuxnet malware that connected via HTTP to two command and control servers to download the infected Windows executables for the engineering workstation. In the initial communication to the command and control servers Stuxnet also sends the IP address of the infected host and a payload consisting of, in part, information on the host OS, the host computer name and domain name, and a flag indicating if the software running on the engineering workstation is Siemens Step7 or the WinCC SCADA environment \cite{xue2018development}. Upon infection the engineering workstation will upload the malicious control logic to the Siemens PLC.
\item \textit{Uncommon Used Port}, used by adversaries for command and control communication to external malicious servers. Such techniques are applicable for firewalls that have been improperly configured as well as can bypass proxies. An example of such technique was used by Crashoverride (i.e. Industroyer) \cite{geiger2020analysis} to communicate to an internal proxy listening on tcp/3128. 
\end{itemize}

A characteristic example of using Command and Control techniques is the Triton malware \cite{di2018triton} that identified the Schneider Electric’s Safety Instrumented Systems (SIS) controller using a network scan and connected to its 1502 port. This port is not common for EPES systems and hence is not included in Table \ref{tab:protocols}. Nevertheless, through this port, Triton was able to connect to external servers and download malicious payload that could harm SIS controllers.

\subsection{Execution}
Execution consists of techniques that result in adversary-controlled code running on a local or remote system, device, or other asset \textcolor{red}{and is based on the MITRE ICS ATT\&CK Matrix \textit{TA0104}\footnote{https://attack.mitre.org/tactics/TA0104/} tactic}. Moreover, this tactic may also rely on unknowing end users or the manipulation of device operating modes to run. Adversaries may infect remote targets with programmed executables or malicious project files that operate according to specified behavior and may alter expected device behavior in subtle ways. Commands for execution may also be issued from command-line interfaces, Application Programming Interfaces (APIs), Graphical User Interfaces (GUIs), or other available interfaces. 

Additionally, adversaries can even execute remote scripts that modify the logic of industrial controllers. Techniques that run malicious code may also be paired with techniques from other tactics, such as Defense and Operator Evasion, Discovery and Disruption. In particular, execution techniques include:
 
\begin{itemize}
\item \textit{API Interaction}, used by adversaries to execute commands that affect industrial controllers through dedicated APIs. In industrial environments, APIs are  based in REST architecture and allow interactions with the industrial controllers through web-service protocols, as HTTP, CoAP and MQTT \cite{lekidis2015using}. Example commands that can be executed once an adversary has access to the API concern the modification of device parameters, the update in the device state (i.e., from START to STOP). Unintentional users (e.g., operators, supply chain personnel) may also cause failures or errors in industrial controllers as they are also using APIs for their remote configuration and maintenance. 
\item \textit{Modify Project File}, used by adversaries to infect project files with malicious code. These project files may consist of objects, program organization units, variables such as tags, documentation, and other configurations needed for industrial controllers programs to function. Using built-in functions of the engineering software, adversaries may be able to download an infected program to a PLC in the operating environment, enabling further execution and persistence techniques. Adversaries may export their own code into project files with conditions to execute at specific intervals. Malicious programs allow adversaries control of all aspects of the process enabled by the PLC. Once the project file is downloaded to a PLC, the workstation device may be disconnected with the infected project file still executing.    
\item \textit{Exploitation of Vulnerability}, used by adversaries to execute commands that affect industrial devices. Vulnerability exploitation techniques usually target weak spots of an industrial system that can be linked to vulnerable interfaces, open ports or software/hardware implementation bugs. Specifically, industrial devices usually contain software that is not properly tested by their manufacturer for vulnerabilities as for instance buffer overflows, employed communication interfaces that can be used as entry point to the industrial system by adversaries.
\item \textit{Graphical User Interface}, used by adversaries to gain remote access to the SCADA or the HMI systems and change critical parameters of the industrial process. Adversaries rely on such techniques since they do not require knowledge of industrial protocols nor any experience on industrial systems (e.g. ladder logic, control system operation). Adversaries may either use physical access or remote access via protocols such as VNC on Linux-based and Unix-based operating systems, and RDP on Windows operating systems. An adversary can use this access to execute programs and applications on the target machine. 
\item \textit{Man In the Middle}, used by adversaries to intercept legitimate communications between industrial devices. Industrial protocols usually lack of authentication and encryption mechanisms, as they were designed for resource-constraint devices with no security in mind. Examples of such protocols are Modbus \cite{swales1999open} and the Siemens S7 \cite{hui2018investigating}. 
\item \textit{Modify Control Logic}, used by adversaries to execute malicious code on a local or remote system that interfaces directly with the industrial process remotely. Such techniques are used by adversaries to enable commands/services or disable safety mechanisms on industrial controllers that would allow them as a following step to interrupt the normal operation of the industrial process. To modify the control logic an adversary needs to have remote access on an industrial controller as well as to intercept and interpret the logic that is the controller is operating on. Examples of such techniques include uploading or updating memory blocks in Rockwell PLCs to use adversary-controlled logic that will maintain input/output at their normal values but will also disable safety mechanisms on the PLC, that will allow unstable behavior, such as putting the PLC into $<Program Mode>$. $<Program Mode>$ allows as a further action to issue a $<Stop CPU>$ command that will shut down the PLC and disrupt the physical process operation. 
\item \textit{Command Line Interface}, used by adversaries to interact with systems and execute commands. CLIs provide a means of interacting with computer systems and are a common feature across many types of platforms and devices within control systems environments. Adversaries may also use CLIs to install and run new software, including malicious tools that may be installed over the course of an operation. CLIs are typically accessed locally, but can also be exposed via services, such as SSH, Telnet, and RDP. Commands that are executed in the CLI execute with the current permissions level of the process running the terminal emulator, unless the command specifies a change in permissions context. Additionally, many industrial controllers have CLI interfaces for management purposes. A characteristic example of using CLI techniques was used during the Dragonfly 2.0 attack \cite{shekari2022madiot}, where PowerShell was used to install the Backdoor.Goodor, a Trojan horse that opens a backdoor on the compromised endpoint.  
\item \textit{Change Program State}, used by adversaries to change the state of the current program on an industrial controller. Program state changes may be used to allow another program to take over control or be loaded onto the controller. The impact of such changes is that it can lead industrial controllers to error-prone states, which require a controller recovery for restoring the normal functionality. Adversaries may use such techniques only if they have remote or physical access to the industrial controllers as well as if they have adequate knowledge about the controller (e.g. protocol, commands, states), as each controller is proprietary and based on its manufacturer. The Change Program State technique can be triggered both intentional and unintentional users (i.e. system operators or supply chain personnel). It shall be considered suspicious when not part of a regular maintenance or troubleshooting. A characteristic example of Change Program State techniques is by using the PLC-Blaster malware \cite{spenneberg2016plc} to change the state of Siemens S7 PLC programs \cite{hui2018investigating}.
\item \textit{Scripting}, used by adversaries to bypass any security mechanisms that are used by industrial controllers as well as to execute user-supplied code without the need of an interpreter. Interpreters read and compile part of the source code just before it is executed, as opposed to compilers, which compile each and every line of code to an executable file. Scripting techniques allow adversaries to run their code on any system where the interpreter exists, such as the Python language that is installed as a default in many Linux distributions. This way, adversaries can distribute one package, instead of precompiling executables for many different systems.  One scenario that such techniques are used lies in the industrial equipment software often does not properly validate user inputs or received data to ensure validity. Hence, an adversary may issue invalid data to allow command injections or cross-site scripting. Another scenario that scripting techniques are used is for disclosing the credentials or cookies of industrial devices. In addition to being a useful tool for developers and administrators, scripting language interpreters may be abused by the adversary to execute code in the target environment. Due to the nature of scripting languages, this allows for weaponized code to be deployed to a target easily, and leaves open the possibility of on-the-fly scripting to perform a task.
\end{itemize}

A characteristic example of using Execution techniques is through the PLC-Blaster software that is a piece of proof-of-concept malware that runs on Siemens S7 PLCs. This worm locates other Siemens S7 PLCs on the network and attempts to infect them by executing malicious code that modifies their control logic. Once this worm has infected its target and attempted to infect other devices on the network, the worm can then run one of many modules. Another example is the use of Ettercap software \cite{pingle2018real} to execute MITM attacks against Workstation and PLCs. During the execution of a MITM attack, an adversary can execute scripts that modify the normal functionality of the EPES system. 

\subsection{Disruption}
Disruption techniques are employed by adversaries to stop normal operation of EPES devices \textcolor{red}{and are based on the MITRE ICS ATT\&CK Matrix \textit{TA0106}\footnote{https://attack.mitre.org/tactics/TA0106/} (i.e., Impair Process Control) and \textit{TA0105}\footnote{https://attack.mitre.org/tactics/TA0105/} (i.e., Impact) tactics}. Such commands are dangerous as may cause catastrophic damages on the industrial control process. Targets of interest may include active procedures or parameters that manipulate the physical environment. These techniques can also include prevention or manipulation of reporting elements and control logic. If adversaries modify process functionality, then they may also obfuscate the results, which are often self-revealing in their impact on the outcome of a product or the environment. As disruption techniques are impacting physical control, they may also threaten the safety of operators and downstream users, which can prompt response mechanisms. 

In this category of techniques there are a lot of methods including changes in the firmware, configuration, inputs/outputs, parameters and messages of industrial controllers that are directly connected to the industrial control process. More sophisticated techniques also aim at changing the memory locations of industrial controllers.  Finally, adversaries can apart from disrupting the operation of industrial controllers and causing Denial of Service (DoS) conditions, adversaries can also stop the existing communications of EPES devices, hence causing Network DoS. In particular, disruption techniques include: 

\begin{itemize}
\item \textit{Alternate Modes Of Operation}, used to modify the operating mode of industrial controllers for causing errors or faults in the industrial process. PLCs have different operating modes that can be accessed by physical, serial but also remote connections. Hence, each PLC manufacturer implements a set of proprietary commands to allow remote control. A characteristic example A characteristic example is with the improper validation of S7 communication packets, which could cause a cold restart of the CPU of Siemens PLCs. The PLC cold restart command for Siemens PLCs that discards all current data and starts program processing again with the initial values that has a direct impact as the existing values of the industrial process are lost. The alternative modes of operation technique can be triggered both intentional and unintentional users (i.e., system operators or supply chain personnel). It shall be considered suspicious when not part of a regular maintenance or troubleshooting. 
\item  \textit{Block Communication Port}, used by adversaries for interrupting the communication between industrial devices, such as to PLC or PLC to Workstation or Human Machine Interface (HMI). Adversaries use such techniques to disrupt interdependent sequential processes, i.e., the processes where one's output is directly used as input to another. Such setup is used in industrial sectors either by using industrial protocols (e.g. Profinet \cite{feld2004profinet}, Profibus \cite{tovar1999real}) or rarely by serial hardwired connections.  The target of the adversary is to cause malfunctions and faulty operations, by blocking the communication for providing inputs to the independent process. Apart from malicious threat actors, blocking of communication ports can be also performed accidentally by i.e. accidental misconfiguration of the firewall or a network settings update by unintentional actors as technicians, operators or even supply chain personnel. A characteristic example of using such techniques is with the Crashoverride malware \cite{geiger2020analysis}, that blocks serial ports of Windows devices, preventing communications between legitimate devices and affected devices.
\item \textit{Block Command Message}, used by adversaries to disrupt the normal operation of the industrial process by blocking legitimate command messages that are intended for industrial controllers. Command messages are used to instruct industrial controllers on performing actions on the industrial process. These actions are often critical, such as changing the speed of rotors for power production systems. An adversary may intercept and block these messages from being delivered to the industrial controllers. Accordingly, industrial controllers will not perform the action that they would do if they properly received the command message. The consequences of blocking command message techniques depend on the criticality of the action associated of the command message for the industrial process. For example, in a bottle filling factory if the command message that stops the water inflow is blocked, the water will overflow from the bottle. On the other hand, a more critical command message that can be blocked is the command for opening or closing the circuit breakers from transmission lines of substations in order to cause unbalance power loads that will have catastrophic failures in the electricity grid of an entire area. This can be accomplished by continuously sending opening commands to numerous breakers, so that if the operator tries to close them back again from the HMI, all requests are overwritten and the breakers remain open, disobeying the orders of the operator from the HMI. To use block command message techniques, adversaries must be present in the industrial network as well as should have adequate knowledge of the industrial network and equipment by using discovery techniques. A characteristic example of using such techniques is with the Crashoverride (i.e. Industroyer) malware \cite{geiger2020analysis} that blocked the control of industrial controllers through the HMI.
\item \textit{Block Reporting Message}, used by adversaries to intercept the diagnostic and reporting messages that are exchanged in industrial systems. Reporting messages are transmitted from industrial controllers to SCADA systems, in order to provide an overview of the state of the monitored system. An adversary may intercept and stop these messages from being received by the SCADA systems that will reduce significantly and often eliminate visibility for the faults or errors that are present in the industrial process. Since the operator will not take action based on the triggered faults, the industrial process may stop working, which will have catastrophic consequences for the industrial process.
\item \textit{Block Serial Communication Port}, used by adversaries to block access to serial COM for preventing instructions or configurations from reaching target devices. Serial Communication ports (COM) allow communication with control system devices. Devices can receive command and configuration messages over such serial COM. Devices also use serial COM to send command and reporting messages. Blocking device serial COM may also block command messages and block reporting messages. A characteristic example is with the Industroyer malware, in which the first COM port from the configuration file is used for the actual communication and the two other COM ports are just opened to prevent other processes accessing them. Thus, the IEC 101 payload component is able to take over and maintain control of the RTU device.
\item \textit{Device Shutdown}, which employ commands that, when issued, interrupt the stop normal operation of industrial devices (e.g. PLCs, RTUs, safety controllers). Such commands are dangerous independently of the industrial process that it is controlling. Each device vendor has a proprietary Stop CPU, hence the adversary must have adequate system knowledge to issue it. The command can also be triggered unintentionally by non-malicious actors, as system operators or supply chain personnel. A characteristic example of such technique is found on the TRISIS malware \cite{mekdad2021threat} that shutdown Triconex controllers of a Saudi petrochemical plant. 
\item \textit{Modify Command Message}, used by adversaries to disrupt the normal operation of the industrial process through the use of dangerous command messages for industrial controllers. Command messages are used to instruct industrial controllers on performing actions on the industrial process. These actions are often critical, such as changing the speed of rotors for power production systems. An adversary may intercept and modify these messages to make the industrial controller perform a different action than the one that it is normally intended to do. This may lead to catastrophic consequences for the industrial process, as a shutdown of an entire production plant. To use modify command message techniques, adversaries must be present in the industrial network as well as should have adequate knowledge of the industrial network and equipment by using discovery techniques.
\item \textit{Modify Control Logic}, used by adversaries to alter the industrial process remotely. To accomplish that, they aim at reprogramming logic of PLCs. Reprogramming will lead the PLC to faulty or wrong reaction, even though the data that is supplied is correct. As a result, unsafe conditions for the industrial process will be allowed, which may as well lead to a catastrophic failure. To use modify control logic techniques, adversaries must be present in the industrial network as well as should have adequate knowledge of the industrial network and equipment by using discovery techniques.
\item \textit{Modify Firmware}, used by adversaries when they issue malicious firmware updates to obtain control in an industrial system or introduce a security weakness that can be leveraged to disrupt the normal system behavior. Firmware is software that is loaded and executed from non-volatile memory on industrial devices in order to initialize and manage device functionality. To use modify firmware techniques, adversaries must be present in the industrial network as well as should have been able to intercept the firmware version that is used in the industrial controllers by using discovery techniques. Additionally, they should be having adequate knowledge of the industrial equipment to update the firmware with malicious commands. Modify firmware can also be issued by unintentional actors (e.g. operators, supply chain personnel) during the controller during maintenance operations. Hence, the distinction between intentional malicious and unintentional actors requires additional indicators from other tactics that are used prior to the disruption tactics.
\item \textit{Modify Physical Device Display}, which aim at manipulating the operational display of industrial systems in order to deceive operators into taking wrong actions that will disrupt the industrial system functionality. Physical Device Display systems are used to display characteristics of the industrial process (e.g. voltage, current in power grids) as well as to monitor the industrial processes as well as the industrial controllers. Examples of such systems are the SCADA or HMI systems. An adversary uses these techniques to deceive operators by displaying faulty or error-prone behaviors for the industrial process. Operators may then trigger fail-safe mechanisms that are installed in order to prevent any physical damage. Certain industrial controller manufacturers (i.e. Siemens) are also using proprietary functionalities to automatically switch to fail-safe mode. In fail-safe mode, the industrial system operates with much less productivity than expected, which causes disruption in the industrial process. Modify Physical Device Display techniques may also be triggered by unintentional users (i.e. operators), when performing maintenance. Hence, the distinction between intentional malicious and unintentional actors requires additional indicators from other tactics that are used prior to the disruption tactics.
\item  \textit{Modify Reporting Message}, used by adversaries to alter the diagnostic and reporting messages that are exchanged in industrial systems. Reporting messages are transmitted from industrial controllers to SCADA systems, in order to provide an overview of the state of the monitored system. An adversary may intercept and modify these messages to make an operator think that an error or fault is present on the system. The operator will then be misguided into taking false actions to restore the system functionality.
\item  \textit{Modify Tag}, which aim at manipulating the industrial process by simultaneously remaining undetected by security controls (network segregation, firewalls, IDS/IPS, etc.). Tag manipulation requires extensive knowledge about the system, that can be achieved by eavesdropping and discovery tactics (i.e., control process and I/O module enumeration). To use modify tag techniques, adversaries must be present in the industrial network as well as should have adequate knowledge of the industrial network and equipment by using discovery techniques. Additionally, tags can also be modified by unintentional actors (e.g. operators, supply chain personnel) without raising any suspicion for security controls that are in place. Hence, the distinction between intentional malicious and unintentional actors requires additional indicators from other tactics that are used prior to the disruption tactics.
\item \textit{Modify System Settings}, aim at manipulating the configuration of an industrial environment. Adversaries use these techniques in, especially, legacy environments where there is no visibility of the system architecture or configuration. The absence of visibility leads system owners to consider ICS devices as "black-boxes", and dread making changes to them. When using such techniques, adversaries modify the connection between the devices or the network architecture (e.g., ports, addressing schemes) to disrupt the normal system functionality. To perform these techniques, an adversary requires a network discovery phase by, for instance performing a man-in-the-middle. The intercepted configuration needs to be carefully modified in order to be accepted by the corresponding industrial devices. Threat origin for these techniques can also be unintentional actors as operators or supply chain personnel when performing maintenance or system updates. 
\item  \textit{Network DoS}, used by adversaries to interrupt the normal functionality of the industrial network. The functionality of the industrial network is to support data and command exchange between the industrial equipment. However, since the industrial communication mechanisms are not standardized, the usual scenario is that the industrial protocol is flexible and each manufacturer can extend it and implement new mechanisms. Usually though, some mechanisms are not compatible with the available resources for the industrial equipment (e.g., memory, processing power, allowed network connections). Adversaries can leverage these incompatibilities, in order to cause denial of service in the network level that stops every existing communication in the industrial system. Stopping the existing communications may result in loosing important network commands at the industrial controller level. The consequences of loosing important network commands are the same as with the Block Command Message techniques, i.e., depend on the criticality of the action associated of the command message for the industrial process. To use Network Denial of Service techniques, adversaries must be present in the industrial network as well as should have adequate knowledge of the industrial network and equipment by using discovery techniques. 
\item \textit{Manipulate I/O Image}, used by adversaries to manipulate the I/O image of PLCs through various means to prevent them from functioning as expected. Methods of I/O image manipulation may include overriding the I/O table via direct memory manipulation or using the override function used for testing PLC programs. During the PLC scan cycle, the state of the actual physical inputs is copied to a portion of the PLC memory, commonly called the input image table. When the program is scanned, it examines the input image table to read the state of a physical input. When the logic determines the state of a physical output, it writes to a portion of the PLC memory, commonly called the output image table. The output image may also be examined during the program scan. To update the physical outputs, the output image table contents are copied to the physical outputs after the program is scanned. One of the unique characteristics of PLCs is their ability to override the status of a physical discrete input or to override the logic driving a physical output coil and force the output to a desired status. 
\item \textit{Spoofing}, used by adversaries to obtain a legitimate identity of personnel affiliated with the industrial system to either perform legitimate actions or conduct DDoS attacks, cyber espionage and information destruction attacks. For instance, adversaries may modify message parts that pose a significant risk and can disrupt the associated industrial process. The difference with spoofing techniques is that adversaries require adequate knowledge of the industrial system through eavesdropping and network discovery. To obtain legitimate identity, they may also use stolen credentials or legitimate authentication tags to leave no suspicious activity traces for security mechanisms as well as forensics. Characteristic examples of masquerading techniques are found in engineering workstations that are running proprietary Windows applications to program industrial controllers.
\item  \textit{Firmware Update Mode}, used by adversaries to cause malfunction or denial of service to industrial controllers. Firmware updates are scheduled for maintenance of the controllers, during which their normal operation is halted. Specifically, till the new firmware is installed and loaded the industrial controller may be placed in an inactive holding state, suspending its usual functionalities. Adversaries may take advantage of this situation to cause denial of service in industrial controllers.   
\item  \textit{Control Device DoS}, used by adversaries to cause denial of service (DoS) condition on PLCs, If adversaries manage to take control of the PLC and trigger fatal errors, this will have a direct impact on the normal operation of the physical process. For triggering DoS conditions, an adversary needs to only have remote access to the PLC, and no specific knowledge of the actual process controlled by the PLC or the program running on it is required. A characteristic example of using such techniques is spotted in Rockwell PLCs, where an unauthenticated user can craft a malicious ACK TCP packet that will immediately reboot the device. After the reboot,  the  device enters a "Major Fault" mode,  which  prevents  normal  operation  of  the  main safety controller program. This mode will not be automatically resolved and requires manual operations to be done by the operator. Another example is with the Rockwell MicroLogix 1400 PLC's is the ability of an unauthenticated user to issue "Change Mode to Program" commands that will stop the running PLC.
\item  \textit{Modify Parameter}, which aim at manipulating parameters of industrial controllers, in order to change the industrial process behavior. Parameters of industrial process are used for configuration of inputs/outputs, for the controller's functionality or even for the communication interfaces. Manipulating parameters can be performed by either intentional malicious or unintentional threat actors (i.e. operators, supply chain personnel). In both cases, threat actors require adequate knowledge of the semantics of the controlled process. A characteristic example of an attack using this technique is the Aurora incident that managed to damage the generator by opening and closing the breakers to lead them in an out-of-sync state. To use modify parameter techniques, adversaries must be present in the industrial network as well as should have adequate knowledge of the industrial network and equipment by using discovery techniques. Modify Parameter techniques may also be triggered by unintentional users (i.e. operators), when performing maintenance. Hence, the distinction between intentional malicious and unintentional actors requires additional indicators from other tactics that are used prior to the disruption tactics.       
\end{itemize}

A characteristic example of using Disruption techniques is the Stuxnet malware that destroyed 1000 centrifuges in Natanz, Iran, by modifying the control logic of Siemens PLCs. Specifically, the modification would turn the rotor of centrifuges in higher speeds that would eventually destroy them.

Using the TTPs that were presented in this section, an adversary can interrupt the normal functionality of EPES systems. In the following section, we present how all these TTPs are mapped into real attack scenarios on different part of the EPES systems.

%% file: caseStudy.tex
\section{Case-study: Cyber-attack in an industrial power plant} \label{sec:caseStudy}

\begin{figure}[htbp!]
\centering
\includegraphics[width=.8\textwidth]{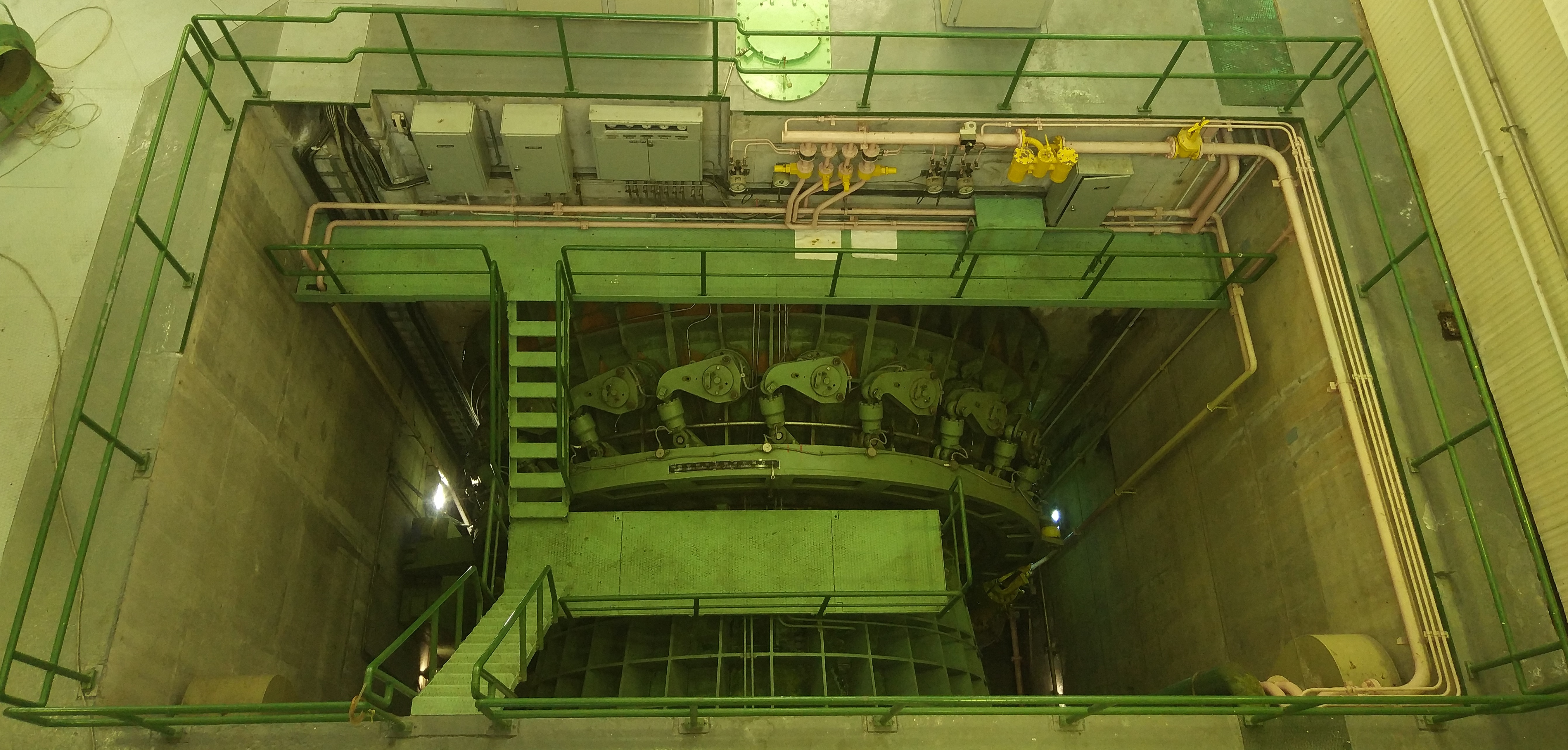}
\caption{Power generator for electricity production}
\label{fig:PowerGenerator}
\end{figure}

In this section, we demonstrate two cyber-attack scenarios in the traditional electricity grid as well as in a smart grid billing scenario. The former aims to cause an operational issue in a power generator of a utility company responsible for electricity production in the Greek area (Figure \ref{fig:PowerGenerator}) and the latter aims to initiate a false data injection attack against a utility's smart meters that are present in distribution substations.

\subsection{Power generator process attack scenario}

The power generator is controlled by PLCs, that send commands to start, stop or reset its operation. The entire system is monitored through the SCADA system that is present in the power plant.

The main objective of this scenario is the disruption of the power generator that is used to produce electrical power on \textcolor{red}{a Hydro Power Plant (HPP)}. This will leave residential areas (e.g., cities) in Greece in blackout that has catastrophic impact for the society. The power generator can be stopped using three actions: 1) issuing a start/stop/reset command from the PLC controlling the power generator (\textit{Device Shutdown} TTP \textcolor{red}{in the \textit{Disruption} tactic}, 2) creating fake incidents on the SCADA to lead the operators into thinking that the power production process is error-prone and hence stop it (\textit{Modify Reporting Message} TTP \textcolor{red}{in the \textit{Disruption} tactic}, 3) setting the temperature of the power generator outside the allowed limits (\textit{Modify Command Message} TTP \textcolor{red}{in the \textit{Disruption} tactic)}.

\begin{figure*}[htbp!]
\includegraphics[width=.67\textwidth]{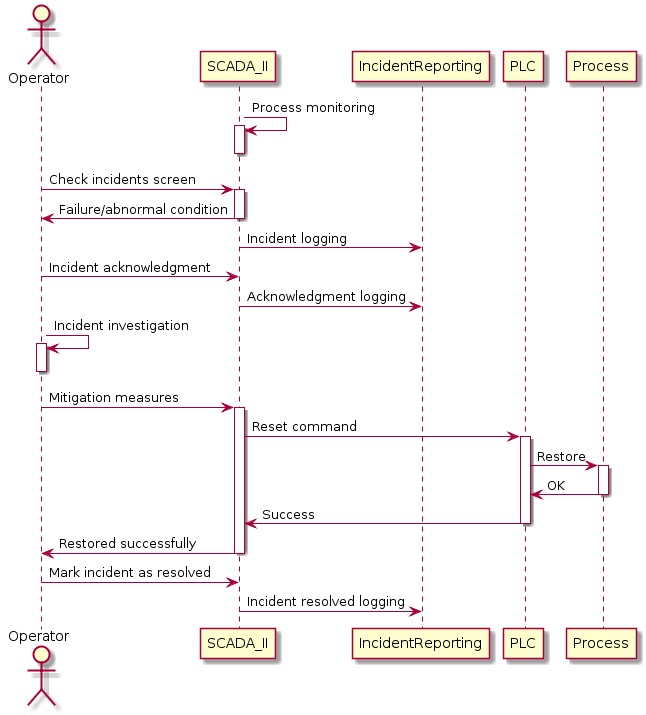}
\centering
\caption{Power generator process attack scenario}
\label{fig:PLCcyber}
\end{figure*}

In order to trigger a cyber-attack that stops the power generator though, an adversary should first obtain access on the electrical production network of the power plant. Then, through the lack of authentication for connecting to the SCADA and then the PLC, an adversary can issue \textit{Device Shutdown} commands to stop the electrical power production of the power plant. To further demonstrate the actions and actors of this scenario we also provide a Unified Modeling Language (UML) diagram in Figure \ref{fig:PLCcyber}. 

The following actors are involved in the Figure:

\begin{itemize}
\item The Operator that is supervising the correct system functionality.
\item The Adversary that is performing malicious actions on the HPP. 
\item The secondary SCADA system (SCADA\_II), supervising the electrical components of the HPP.
\item The PLC, representing the field controllers that are interfacing with the power generator.
\item The Power Generator (PowerGenerator), that controls the power production in the HPP. 
\end{itemize}

Based on these actors, the steps of the attack scenario are identified as follows:
\begin{enumerate}
\item The adversary establishes a remote Telnet connection with the SCADA (SCADA\_II).
\item The adversary verifies that access on the PLCs is obtained. 
\item The adversary schedules a power generator halt command from the SCADA II.
\item The SCADA (SCADA\_II) sends the command to the PLCs.
\item The PLCs stop the power generator operation. 
\item The SCADA (SCADA\_II) informs the Operator that the production is stopped due to an error (i.e. power generator halt). 
\item The Operator starts the incident investigation in order acknowledge and respond to the occurred incident. 
\end{enumerate}

 \begin{figure*}[htbp!]
\centering
\includegraphics[width=1.1\textwidth]{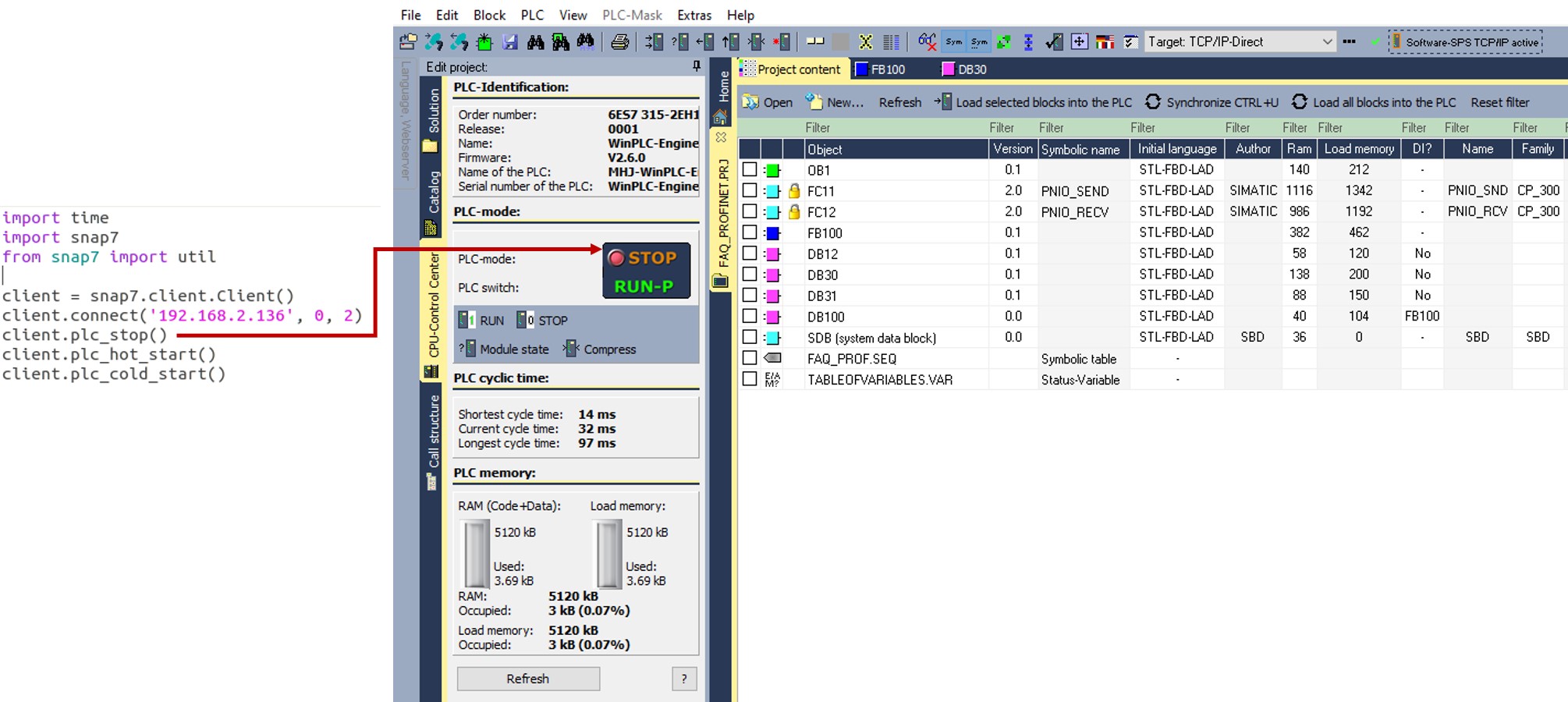}
\caption{Stopping the PLC of an EPES power plant}
\label{fig:attackDemonstration}
\end{figure*}

Interfering with the PLC operation is accomplished using the Snap7 Python library\footnote{https://pypi.org/project/python-snap7/} to connect to the Siemens PLCs. Then, an appropriate command is transmitted to the that halts the PLC operation and subsequently the power generator as well (Figure \ref{fig:attackDemonstration}).

\subsection{Smart meter attack scenario}

In this scenario the adversary is accessing the data exchanged from the smart meters towards the distribution control center by initially relying on initial access, which is performed by eavesdropping the wireless connection between the smart meters and the mobile base station. \textcolor{red}{Then, the adversary is using \textit{Network Sniffing} TTP that resides in the Discovery tactic. The smart meter that is considered in this attack is Landis+Gyr E650\footnote{https://www.landisgyr.eu/product/landisgyr-e650/} and the conducted MITM attacks were performed through the use of the Ettercap software \cite{norton2004ettercap}. Ettercap allows to perform Address Resolution Protocol (ARP) spoofing and hence we managed to intercept the AARQ request, which provided smart meter measurements to the adversarial device}. Then, the smart meter measurements can be altered (tampering attack) \textcolor{red}{using the \textit{Disruption} tactic and specifically the \textit{Modify Parameter} TTP, or the attacker can also steal the identity using \textit{Credential access} tactics for trying to connect to the HPP SCADA. }

\begin{figure*}[htbp!]
\centering
\includegraphics[width=1\textwidth]{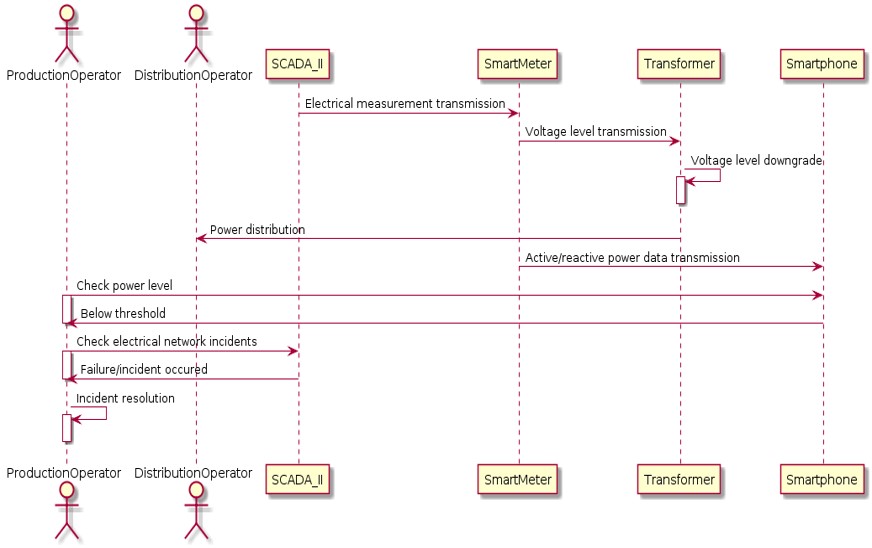}
\caption{Smart meter attack scenario}
\label{fig:smCyber}
\end{figure*}

To demonstrate the actions and actors of this scenario we also provide a UML diagram in Figure \ref{fig:smCyber}. As illustrated by the figure the main actors for this scenario are:
\begin{itemize}
\item The ProductionOperator that is responsible for maintaining the proper electrical network functionality on the production side.
\item The DistributionOperator that is responsible for the distribution of the produced power from the HPPs. 
\item The SCADA system (SCADA\_II), supervising the industrial components of the electrical network.
\item The SmartMeter, that collects the electrical measurements from the product site and transmits the voltage levels to the transformers.
\item The Transformer that downgrades the voltage level to medium voltage such that can be distributed by the DistributionOperator. 
\item The Smartphone that is logging the active/reactive power data which is then made available for power level monitoring to the ProductionOperator.
\end{itemize}

Based on these actors the attack scenario that is followed consists of the steps below:

\begin{enumerate}
\item Eavesdropping of wireless connections while being physically close to the utility substation. 
\item Perform \textcolor{red}{MITM} attack near the smart meter to disrupt the established session communication.  
\item Alter the active and reactive power measurements in the exchanged packets \textcolor{red}{in DLMS/COSEM \cite{lekidis2024towards} protocol format}.
\item The false measurements are logged in the Utility Data Control Center, which calculates the overall electricity consumption and billing based on the received measurements. 
\item Use Lateral Movement techniques to manipulate measurements of all nearby smart meters or the SCADA\_II system in order to cause instabilities to the electrical grid.
\item Incident is logged in the SCADA\_II system and the ProductionOperator or DistributionOperator is informed to initiate mitigation procedures.
\end{enumerate}

\begin{figure}[htbp!]
    \centering
    \includegraphics[width=.8\linewidth]{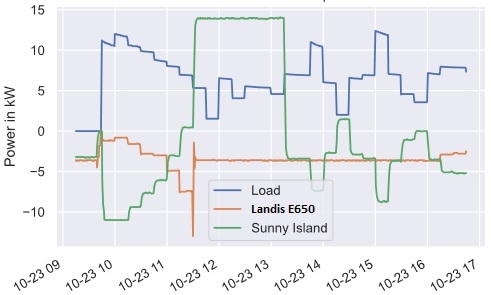}
    \caption{\textcolor{red}{{Smart meter attack (based on \cite{lekidis2023risk}}}}
    \label{fig:smartMeterAttack}
\end{figure}

\textcolor{red}{As an outcome of the conducted attack scenario the measurements that are produced by the smart meters are presented in Figure \ref{fig:smartMeterAttack}. Specifically, this figure depicts a data tampering attack falsifying the measurements of the Landis+Gyr E650 smart meter (orange line in Figure \ref{fig:smartMeterAttack}), to make them negative in the scope of the attack scenario and cause the below threshold message of Figure \ref{fig:smCyber}. The actual load of the scenario is depicted in the blue line of Figure \ref{fig:smartMeterAttack} and also the figure includes a Sunny Island inverter (green line) which is used for facilitating energy storage in a battery system.}

%% file: discussion.tex
\section{Discussion} \label{sec:discussion}

\textcolor{red}{In this section we detail the lessons learned from the conducted cyber-attack scenarios of Section \ref{sec:caseStudy} to validate the proposed method, provide detection and mitigation strategies for each one of the demonstrated attack scenarios as well as describe the limitations of the proposed method. }

\textcolor{red}{\subsection{Lessons Learned}}
Upon applying the identified TTPs in a real EPES power plant infrastructure, we have identified lessons learned that are presented for both power production halt and smart meter attack scenarios. Initially, the PLC and SCADA systems include no update/upgrade mechanisms due to the safety and security
reasons of disallowing external connections. Hence, the associated EPES devices are prone to faults or
failures as they contain legacy software that a) may contain bugs or errors, b) not fully supported by its vendor. Furthermore, when faults or failures
occur there are very few incident response procedures and the personnel that is aware of maintenance
and programming operations is limited. For certain devices, maintenance is entirely assigned to third-parties and no utility personnel is involved in maintenance and programming operations.

Additionally, the communication between all the power plant assets is unencrypted and there is no access
control nor any authentication available on the internal communication network. Combined with the
limited physical security controls to prevent unauthorized personnel from entering the power plant area during
working hours, this creates a very serious threat for the power plant operation. Specifically, an adversary
can enter the HPP area and use social engineering means to locate the electrical network. Once having
identified the electrical network, the next step is to connect a malicious device on the internal
network. This device may eavesdrop the internal communication but also send malicious commands
that will be executed by PLCs and hence cause disruption on the electrical generator. The criticality of
these vulnerabilities is increased if we consider that though there are backup EPES devices to ensure
redundancy, not all are properly installed nor configured.

Concerning the smart meter attack scenario, the smart meters often use cellular connections (i.e., 3G) with no
authentication nor authorization to communicate with the distribution network operator. The smart
meters are used for logging non-critical measurements as the voltage, current and active/reactive
power. Even though these measurements can be tampered leading to electricity theft as in the case of
Enemalta \cite{lekidis2022cyber}, this is not the most critical vulnerability that was discovered from our assessment. An
adversary can eavesdrop the cellular connection and try to connect to the smart meter in order to 1)
trigger a disconnect power operation or 2) move laterally from the substation to the HPP system and
propagate the cyber-attack to more critical EPES devices of the power plant (i.e., PLCs, SCADA system).

\textcolor{red}{\subsection{Mitigation mechanisms for cyber-resilience}} \label{sec:mitigations}

\textcolor{red}{The conducted attacks in Section \ref{sec:caseStudy} indicate the need for mitigation measures to ensure cyber-resilience. }
\begin{table}[htbp!]
  \centering
    \resizebox{9.5cm}{!} {
    \begin{tabular} {|c||c|}
        \hline                                                                
\textbf{Attack steps}   & \textbf{Potential detection methods}         \\                                                                 \hline                                           
\makecell{Leverage inadequate physical \\  security measures to access \\ SCADA station}                                                                                        & \makecell{Events monitoring, \\ intrusion   detection}      \\                                                                             \hline 
\makecell{Issue a start/ stop/ reset \\  command to the PLC controlling the \\ power generator}                                                                                   & \makecell{Configuration changes, \\ detect   abnormal behavior}    \\                                                                     \hline  
\makecell{Change the network configuration \\  to cause fatal errors in \\ the system operation}                                                                                  & \makecell{Events monitoring, \\ monitor process execution,\\ detect anomalies, \\ intrusion detection, \\ detect abnormal   behavior}  \\       \hline 
\makecell{Set the temperature of the power \\  generator off limits }                            & \makecell{Events monitoring, monitor \\  execution process, detect anomalies, \\ intrusion detection} \\         \hline  
\makecell{Power production stops}  &  \makecell{Events monitoring, monitor  \\ execution process, detect anomalies,\\ intrusion detection}                                 \\    \hline  
\makecell{Catastrophic damages in the power \\  generator that stop power \\ generator until it is replaced}                                                                      & \makecell{Configuration changes, monitor \\  execution process, \\ intrusion detection}                                              \\      \hline 
\end{tabular}}
\vspace{.2cm}
\caption{\textcolor{red}{Detection techniques for the industrial power plant attack scenario}} 
  \label{tab:mitigationsSCADA}%
  \end{table}
\textcolor{red}{Hence, for each attack scenario we have conducted an analysis of the methods that can be used for their detection and mitigation. The analysis was conducted per attack scenario and hence Table \ref{tab:mitigationsSCADA} presents the industrial power plant attack scenario.}

\textcolor{red}{Then, Table \ref{tab:smartMeterAttack} presents the detection mechanisms that can be added to the smart meter attack scenario.}

\begin{table}[htbp!]
  \centering
    \resizebox{9cm}{!} {
    \begin{tabular} {|c||c|}
        \hline                                                                
\textbf{Attack steps}   & \textbf{Potential detection methods}         \\                                                                 \hline                                           
\makecell{Gain access to \\ cellular connection}                                                                                        & \makecell{Events monitoring, \\ intrusion   detection}      \\                                                                             \hline 
\makecell{Connect to the IP address \\ of the smart meter}                                                                                   & \makecell{Configuration management, \\ detect a typical logins and \\ entity behavior, intrusion \\ detection, two factor \\ authentication}    \\                                                                     \hline  
\makecell{Decrypt the \\ smart meter messages \\ (i.e., in DLMS/COSEM format)}                                                                                  & \makecell{Events monitoring, \\ intrusion   detection}  \\       \hline 
\makecell{Alter the exchanged \\ active/reactive power data}                            & \makecell{Events monitoring, \\ monitor process execution, \\ detect anomalies, \\ intrusion detection, \\ detect abnormal behavior} \\         \hline  
\end{tabular}}
\vspace{.2cm}
\caption{\textcolor{red}{Detection techniques for the smart meter attack scenario}} 
  \label{tab:smartMeterAttack}%
  \end{table}

\textcolor{red}{Furthermore, aside from these detection mechanisms, there are also mitigation mechanisms that are applied to the attack scenarios of Section \ref{sec:caseStudy} that are based on the following measures:}

\begin{enumerate}
    \item \textcolor{red}{PLC reset allowing to restore the factory settings and configurations on the PLC and allow it to return to its nominal operation (control of the industrial power plant) after the attack occurrence. 
    \item Restart Windows station which applies to both the Workstation and the SCADA systems to initiate their working sequence.
    \item Activate manual HPP control which applies to both the PLC and SCADA systems to maintain nominal system operation for the industrial power plant attack scenario. 
    \item Isolate infected devices which applies to the PLC, Workstation and SCADA systems of the industrial power plant attack scenario and can be possible by for instance creating a Virtual Local Area Network (VLAN) where the infected devices will be placed. 
    \item Block the access of compromised or malicious devices or systems that the adversary uses to perform the industrial power plant or the smart meter attack scenario. This measure applies to PLC, Workstation, SCADA as well as smart meters and is possible for instance through the use of a firewall mechanism.
    \item Block the access of an compromised or malicious process or application on critical systems as the industrial power plant. This measure applies to both the Workstation and SCADA systems and the processes as well as applications they are executing.
    \item Replace a compromised with a backup device to ensure redundancy and maintain the business operation of both the industrial power plant and the smart meter attack scenarios. Hence, this measure applies to PLC, Workstation, SCADA as well as smart meters.
    \item Use encrypted communication, which aids in avoiding eavesdropping and MITM attacks that are applicable mainly to the smart meter attack scenario. As an example a mutual identification in DLMS/COSEM is followed in the High-Level Security (HLS) \cite{lekidis2023risk}.
}

\textcolor{red}{These measures aid in mitigating the industrial power plant and smart meter attack scenarios, however a concrete and active incident response plan, as in \cite{lekidis2024towards} for the smart meters, should be in place to ensure the availability and continuous operation of such EPES systems. }
\end{enumerate}

\textcolor{red}{\subsection{EPES TTP approach limitations}}

\textcolor{red}{Although the proposed approach is trying to interpret the TTPs of EPES systems, it cannot also cover the Artificial Intelligence (AI) models that are employed in EPES, allowing automation in the processes and procedures that they are using towards the integration in a smart grid architecture \cite{kabalci2019introduction}. Such models are linked to energy demand forecasting \cite{lekidis2023edge}, predictive maintenance of individual components from the different EPES layers in Figure \ref{fig:epesArch} \cite{lekidis2024predictive}. To allow considering such models the proposed method could also use standardized methods as the MITRE ATLAS\footnote{https://atlas.mitre.org/}. MITRE ATLAS would allow the method to cover also adversarial AI attack techniques, which are interrupting the nominal operation of the AI-assisted EPES systems \cite{bountakas2023defense}. }

\textcolor{red}{Additionally, the method would benefit from a validation for the detection and mitigation strategies in the presented TTPs. In the previous part (Section \ref{sec:mitigations}) the detection mechanisms for the attack scenarios in Section \ref{sec:caseStudy} are identified and as well as the produced dataset in \cite{dataset2025} also provides detection mechanisms for the considered TTPs and attack scenarios. Nevertheless, these mechanisms are indicative and their effectiveness still needs to be proved through a real deployment in an EPES system architecture. }

\textcolor{red}{Finally, understanding the TTPs in EPES systems is usually the initial step towards the risk assessment, which provides an overview of the risk posture as well as the criticality in addressing vulnerabilities and attack surfaces \cite{lekidis2023risk}. The current work requires enrichment in order to be used as a reference for prioritizing the risks related to the EPES system infrastructure.}

%% file: conclusion.tex
\section{Conclusion} \label{sec:conc}

This article presents a thorough overview of the cyber-attack steps as well as the TTPs used by cyber-attackers when targeting EPES systems. The TTPs are divided into dedicated categories, and each category includes examples of known attacks and exploited vulnerabilities on the electrical grid by adversaries. The insights gained through the TTPs analysis are applied in an industrial power plant, which includes PLCs to control generator operation as well as smart meters to measure the generated electricity. The lessons learned by applying the TTPs in the power plant aid towards the identification of risks in the power plant nominal operation.

As a part of our future work, we plan to investigate industrial network-based intrusion detection mechanisms as presented in \cite{dupont2019evaluation} for the TTPs that were presented in this article. The mechanisms need to operate under the requirement of zero impact or change in the entire industrial process operation. Moreover, we also will propose a list of mitigation actions as countermeasures for the detected cyber-attacks.

%% file: main.bbl
\begin{thebibliography}{49}
\newcommand{\enquote}[1]{``#1''}
\providecommand{\natexlab}[1]{#1}
\providecommand{\url}[1]{\normalfont{#1}}
\providecommand{\urlprefix}{}

\bibitem[Adamov, Carlsson, and Surmacz(2019)]{adamov2019analysis}
Adamov, Alexander, Anders Carlsson, and Tomasz Surmacz. 2019. ``An analysis of
  lockergoga ransomware.'' In \emph{2019 IEEE East-West Design \& Test
  Symposium (EWDTS)}, 1--5. IEEE.

\bibitem[Adepu and Mathur(2018)]{adepu2018assessing}
Adepu, Sridhar, and Aditya Mathur. 2018. ``Assessing the effectiveness of
  attack detection at a hackfest on industrial control systems.'' \emph{IEEE
  Transactions on Sustainable Computing} 6 (2): 231--244.

\bibitem[Alotaibi and Vassilakis(2021)]{alotaibi2021sdn}
Alotaibi, Fahad~M, and Vassilios~G Vassilakis. 2021. ``Sdn-based detection of
  self-propagating ransomware: the case of badrabbit.'' \emph{IEEE Access} 9:
  28039--28058.

\bibitem[Assante and Lee(2015)]{assante2015industrial}
Assante, Michael~J, and Robert~M Lee. 2015. ``The industrial control system
  cyber kill chain.'' \emph{SANS Institute InfoSec Reading Room} 1.

\bibitem[Axelson(1999)]{axelson1999designing}
Axelson, Jan. 1999. ``Designing RS-485 circuits.'' \emph{Circuit Cellar} 107:
  20--24.

\bibitem[Barai, Krishnan, and Venkatesh(2015)]{barai2015smart}
Barai, Gouri~R, Sridhar Krishnan, and Bala Venkatesh. 2015. ``Smart metering
  and functionalities of smart meters in smart grid-a review.'' In \emph{2015
  IEEE Electrical Power and Energy Conference (EPEC)}, 138--145. IEEE.

\bibitem[Bhamare et~al.(2020)]{bhamare2020cybersecurity}
Bhamare, Deval, Maede Zolanvari, Aiman Erbad, Raj Jain, Khaled Khan, and Nader
  Meskin. 2020. ``Cybersecurity for industrial control systems: A survey.''
  \emph{computers \& security} 89: 101677.

\bibitem[Bountakas et~al.(2023)]{bountakas2023defense}
Bountakas, Panagiotis, Apostolis Zarras, Alexios Lekidis, and Christos Xenakis.
  2023. ``Defense strategies for adversarial machine learning: A survey.''
  \emph{Computer Science Review} 49: 100573.

\bibitem[Brazil(2014)]{brazil2014security}
Brazil, Jody. 2014. ``Security metrics to manage change.'' \emph{Network
  Security} 2014 (10): 5--7.

\bibitem[Butun, Lekidis, and dos Santos(2020)]{butun2020security}
Butun, Ismail, Alexios Lekidis, and Daniel~Ricardo dos Santos. 2020. ``Security
  and Privacy in Smart Grids: Challenges, Current Solutions and Future
  Opportunities.'' \emph{ICISSP} 10: 0009187307330741.

\bibitem[Byres and Lowe(2004)]{byres2004myths}
Byres, Eric, and Justin Lowe. 2004. ``The myths and facts behind cyber security
  risks for industrial control systems.'' In \emph{Proceedings of the VDE
  Kongress}, Vol. 116, 213--218.

\bibitem[Case(2016)]{case2016analysis}
Case, Defense~Use. 2016. ``Analysis of the cyber attack on the Ukrainian power
  grid.'' \emph{Electricity Information Sharing and Analysis Center (E-ISAC)}
  388: 1--29.

\bibitem[Caselli and Kargl(2016)]{caselli2016security}
Caselli, Marco, and Frank Kargl. 2016. ``A security assessment methodology for
  critical infrastructures.'' In \emph{International Conference on Critical
  Information Infrastructures Security}, 332--343. Springer.

\bibitem[Chen and Abu-Nimeh(2011)]{chen2011lessons}
Chen, Thomas~M, and Saeed Abu-Nimeh. 2011. ``Lessons from stuxnet.''
  \emph{Computer} 44 (4): 91--93.

\bibitem[Clarke, Reynders, and Wright(2004)]{clarke2004practical}
Clarke, Gordon, Deon Reynders, and Edwin Wright. 2004. \emph{Practical modern
  SCADA protocols: DNP3, 60870.5 and related systems}. Newnes.

\bibitem[Di~Pinto, Dragoni, and Carcano(2018)]{di2018triton}
Di~Pinto, Alessandro, Younes Dragoni, and Andrea Carcano. 2018. ``TRITON: The
  first ICS cyber attack on safety instrument systems.'' In \emph{Proc. Black
  Hat USA}, Vol. 2018, 1--26.

\bibitem[Dupont et~al.(2019)]{dupont2019evaluation}
Dupont, Guillaume, Jerry Den~Hartog, Sandro Etalle, and Alexios Lekidis. 2019.
  ``Evaluation framework for network intrusion detection systems for in-vehicle
  can.'' In \emph{2019 IEEE International Conference on Connected Vehicles and
  Expo (ICCVE)}, 1--6. IEEE.

\bibitem[Falliere, Murchu, and Chien(2011)]{falliere2011w32}
Falliere, Nicolas, Liam~O Murchu, and Eric Chien. 2011. ``W32. stuxnet
  dossier.'' \emph{White paper, symantec corp., security response} 5 (6): 29.

\bibitem[Fayi(2018)]{fayi2018petya}
Fayi, Sharifah Yaqoub~A. 2018. ``What Petya/NotPetya ransomware is and what its
  remidiations are.'' In \emph{Information Technology-New Generations: 15th
  International Conference on Information Technology}, 93--100. Springer.

\bibitem[Feld(2004)]{feld2004profinet}
Feld, Joachim. 2004. ``PROFINET-scalable factory communication for all
  applications.'' In \emph{IEEE International Workshop on Factory Communication
  Systems, 2004. Proceedings.}, 33--38. IEEE.

\bibitem[Firoozjaei et~al.(2022)]{firoozjaei2022evaluation}
Firoozjaei, Mahdi~Daghmehchi, Nastaran Mahmoudyar, Yaser Baseri, and Ali~A
  Ghorbani. 2022. ``An evaluation framework for industrial control system cyber
  incidents.'' \emph{International Journal of Critical Infrastructure
  Protection} 36: 100487.

\bibitem[Geiger et~al.(2020)]{geiger2020analysis}
Geiger, Marcus, Jochen Bauer, Michael Masuch, and J{\"o}rg Franke. 2020. ``An
  analysis of black energy 3, Crashoverride, and Trisis, three malware
  approaches targeting operational technology systems.'' In \emph{2020 25th
  IEEE International Conference on Emerging Technologies and Factory Automation
  (ETFA)}, Vol.~1, 1537--1543. IEEE.

\bibitem[Hui and McLaughlin(2018)]{hui2018investigating}
Hui, Henry, and Kieran McLaughlin. 2018. ``Investigating current PLC security
  issues regarding Siemens S7 communications and TIA portal.'' In \emph{5th
  International Symposium for ICS \& SCADA Cyber Security Research 2018 5},
  67--73.

\bibitem[Jorquera~Valero et~al.(2022)]{jorquera2022design}
Jorquera~Valero, Jos{\'e}~Mar{\'\i}a, Pedro~Miguel S{\'a}nchez~S{\'a}nchez,
  Alexios Lekidis, Javier Fernandez~Hidalgo, Manuel Gil~P{\'e}rez, M~Shuaib
  Siddiqui, Alberto Huertas~Celdr{\'a}n, and Gregorio Mart{\'\i}nez~P{\'e}rez.
  2022. ``Design of a Security and Trust Framework for 5G Multi-domain
  Scenarios.'' \emph{Journal of Network and Systems Management} 30 (1): 1--35.

\bibitem[Kabalci and Kabalci(2019)]{kabalci2019introduction}
Kabalci, Ersan, and Yasin Kabalci. 2019. ``Introduction to smart grid
  architecture.'' \emph{Smart grids and their communication systems} 3--45.

\bibitem[Kargl et~al.(2014)]{kargl2014insights}
Kargl, Frank, Rens~W van~der Heijden, Hartmut K{\"o}nig, Alfonso Valdes, and
  Marc~C Dacier. 2014. ``Insights on the security and dependability of
  industrial control systems.'' \emph{IEEE security \& privacy} 12 (6): 75--78.

\bibitem[Lekidis(2022)]{lekidis2022cyber}
Lekidis, Alexios. 2022. ``Cyber-security measures for protecting EPES systems
  in the 5G area.'' In \emph{Proceedings of the 17th International Conference
  on Availability, Reliability and Security}, 1--10.

\bibitem[Lekidis(2025)]{dataset2025}
Lekidis, Alexios. 2025. ``Cyber-attack detection, validation and response
  strategies for Electrical Power \& Energy Systems.''
  \urlprefix\url{https://data.mendeley.com/datasets/93st6ddp9r/1}.

\bibitem[Lekidis et~al.(2024)]{lekidis2024predictive}
Lekidis, Alexios, Angelos Georgakis, Christos Dalamagkas, and Elpiniki~I
  Papageorgiou. 2024. ``Predictive maintenance framework for fault detection in
  remote terminal units.'' \emph{Forecasting} 6 (2): 239--265.

\bibitem[Lekidis, Mavroeidis, and Fysarakis(2024)]{lekidis2024towards}
Lekidis, Alexios, Vasileios Mavroeidis, and Konstantinos Fysarakis. 2024.
  ``Towards Incident Response Orchestration and Automation for the Advanced
  Metering Infrastructure.'' In \emph{2024 IEEE 20th International Conference
  on Factory Communication Systems (WFCS)}, 1--8. IEEE.

\bibitem[Lekidis and Papageorgiou(2023{\natexlab{a}})]{lekidis2023edge}
Lekidis, Alexios, and Elpiniki~I Papageorgiou. 2023{\natexlab{a}}. ``Edge-based
  short-term energy demand prediction.'' \emph{Energies} 16 (14): 5435.

\bibitem[Lekidis and Papageorgiou(2023{\natexlab{b}})]{lekidis2023risk}
Lekidis, Alexios, and Elpiniki~I Papageorgiou. 2023{\natexlab{b}}. ``Risk
  assessment method for 5g-oriented dlms/cosem communications.'' In \emph{2023
  IEEE Conference on Standards for Communications and Networking (CSCN)},
  15--21. IEEE.

\bibitem[Lekidis et~al.(2015)]{lekidis2015using}
Lekidis, Alexios, Emmanouela Stachtiari, Panagiotis Katsaros, Marius Bozga, and
  Christos~K Georgiadis. 2015. ``Using BIP to reinforce correctness of
  resource-constrained IoT applications.'' In \emph{10th IEEE International
  Symposium on Industrial Embedded Systems (SIES)}, 1--10. IEEE.

\bibitem[Mackiewicz(2006)]{mackiewicz2006overview}
Mackiewicz, Ralph~E. 2006. ``Overview of IEC 61850 and Benefits.'' In
  \emph{2006 IEEE Power Engineering Society General Meeting}, 8--pp. IEEE.

\bibitem[Matou{\v{s}}ek(2017)]{matouvsek2017description}
Matou{\v{s}}ek, Petr. 2017. ``Description and analysis of IEC 104 Protocol.''
  \emph{Faculty of Information Technology, Brno University o Technology, Tech.
  Rep} .

\bibitem[Mekdad et~al.(2021)]{mekdad2021threat}
Mekdad, Yassine, Giuseppe Bernieri, Mauro Conti, and Abdeslam~El Fergougui.
  2021. ``A threat model method for ICS malware: the TRISIS case.'' In
  \emph{Proceedings of the 18th ACM International Conference on Computing
  Frontiers}, 221--228.

\bibitem[Mohurle and Patil(2017)]{mohurle2017brief}
Mohurle, Savita, and Manisha Patil. 2017. ``A brief study of wannacry threat:
  Ransomware attack 2017.'' \emph{International Journal of Advanced Research in
  Computer Science} 8 (5): 1938--1940.

\bibitem[Norton(2004)]{norton2004ettercap}
Norton, Duane. 2004. ``An ettercap primer.'' \emph{SANS Institute InfoSec
  Reading Room} 5.

\bibitem[Obregon(2015)]{obregon2015secure}
Obregon, Luciana. 2015. ``Secure architecture for industrial control systems.''
  \emph{SANS Institute InfoSec Reading Room} 2.

\bibitem[Pingle, Mairaj, and Javaid(2018)]{pingle2018real}
Pingle, Bhargav, Aakif Mairaj, and Ahmad~Y Javaid. 2018. ``Real-world
  man-in-the-middle (MITM) attack implementation using open source tools for
  instructional use.'' In \emph{2018 IEEE International Conference on
  Electro/Information Technology (EIT)}, 0192--0197. IEEE.

\bibitem[Rascagneres and Lee(2018)]{rascagneres2018wasn}
Rascagneres, Paul, and Martin Lee. 2018. ``Who wasn’t responsible for Olympic
  Destroyer.'' \emph{Talos Intelligence: Talos} .

\bibitem[Sarfi, Green, and Simmins(2012)]{sarfi2012ami}
Sarfi, Robert, Brian~D Green, and John Simmins. 2012. ``AMI network (AMI
  head-end to/from smart meters).'' \emph{Onramp Wireless, San Diego, CA, USA,
  Tech. Rep} .

\bibitem[Shekari, Cardenas, and Beyah(2022)]{shekari2022madiot}
Shekari, Tohid, Alvaro~A Cardenas, and Raheem Beyah. 2022. ``$\{$MaDIoT$\}$
  2.0: Modern $\{$High-Wattage$\}$$\{$IoT$\}$ Botnet Attacks and Defenses.'' In
  \emph{31st USENIX Security Symposium (USENIX Security 22)}, 3539--3556.

\bibitem[Spenneberg, Br{\"u}ggemann, and Schwartke(2016)]{spenneberg2016plc}
Spenneberg, Ralf, Maik Br{\"u}ggemann, and Hendrik Schwartke. 2016.
  ``PLC-blaster: A worm living solely in the PLC.'' \emph{Black Hat Asia} 16:
  1--16.

\bibitem[Stallings(1993)]{stallings1993snmp}
Stallings, William. 1993. \emph{SNMP, SNMPv2, and CMIP: The practical guide to
  network management}. Addison-Wesley Longman Publishing Co., Inc.

\bibitem[Swales et~al.(1999)]{swales1999open}
Swales, Andy, et~al. 1999. ``Open modbus/tcp specification.'' \emph{Schneider
  Electric} 29: 3--19.

\bibitem[Tovar and Vasques(1999)]{tovar1999real}
Tovar, Eduardo, and Francisco Vasques. 1999. ``Real-time fieldbus
  communications using Profibus networks.'' \emph{IEEE transactions on
  Industrial Electronics} 46 (6): 1241--1251.

\bibitem[Tuballa and Abundo(2016)]{tuballa2016review}
Tuballa, Maria~Lorena, and Michael~Lochinvar Abundo. 2016. ``A review of the
  development of Smart Grid technologies.'' \emph{Renewable and Sustainable
  Energy Reviews} 59: 710--725.

\bibitem[Xue et~al.(2018)]{xue2018development}
Xue, Guangku, Yongzhong Qiao, Yuanmeng Xia, Ying Li, Zhi Cheng, Si~Lei, and
  Li~Wang. 2018. ``The development and application of online fault waring
  system for automatic stereo library based with S7-300 and Wincc flexible 2007
  operating environment.'' In \emph{2018 IEEE 9th International Conference on
  Mechanical and Intelligent Manufacturing Technologies (ICMIMT)}, 197--201.
  IEEE.

\end{thebibliography}
